\documentclass[%
reprint,
superscriptaddress,
amsmath,
amssymb,
aps,
pra,
longbibliography
]{revtex4-1}
\usepackage{graphicx}
\usepackage{dcolumn}
\usepackage{bm}
\usepackage{color}
\usepackage[T1]{fontenc}
\usepackage{times}
\usepackage{amsmath,amsfonts,amssymb,mathtools}
\usepackage{hyperref}
\usepackage{bbold}
\usepackage{tensor}
\usepackage{multirow}
\usepackage{diagbox}

\begin{document}
\preprint{APS/123-QED}

\title{Polarization dependent non-Hermitian atomic grating controlled by dipole blockade effect}
\author{Yi-Mou Liu}\email{liuym605@nenu.edu.cn}
 \affiliation{Center for Quantum Sciences and School of Physics, Northeast Normal University, Changchun 130024, P. R. China.}
  \author{Lin Zhang}
 \affiliation{Center for Quantum Sciences and School of Physics, Northeast Normal University, Changchun 130024, P. R. China.}

  \date{\today}

\begin{abstract}
We propose a theoretical scheme for a non-Hermitian atomic grating within a ultra-cold rubidium-87 ($^{87}Rb$) atomic ensemble. The grating's diffraction properties depend on the polarization states of incident photons and are controlled non-locally through Rydberg interactions. Multiple types of polarization-dependent diffraction modes are generated, benefiting from no crosstalk atomic transition channels based on transition selection rules. Those polarization-dependent diffraction modes can be switched using dynamic optical pulse trains, exploiting the Rydberg blockade effect, and are tunable by non-Hermitian optical modulation. Our work will advance the application of asymmetric optical scattering by utilizing the polarization degree of freedom within continuous media and benefit the application of versatile non-Hermitian/asymmetric optical devices.
\end{abstract}
\maketitle

\section{Introduction}

As a fundamental optical device, the grating operates on the principle that optical media in spectroscopic systems have distinct refractive indices for different light frequencies. As a result, light beams acquire corresponding diffraction angle spectra depending on frequencies after being far-field diffracted and coherently superimposed. With advancements in current technology, the performance indicators of gratings, such as reflectivity, diffraction efficiency, and aberration, have significantly improved and advanced. Combined with binary optics algorithms~\cite{J2021, D2022, C2024}, it has become possible to design and process gratings with a wide range of diffraction properties to suit the needs of many modern technologies, including optical detection ~\cite{OD2010-01, OD2009-02, OD2014-03, OD2005-04, OD1996-05}, spectroscopy~\cite{S2015-01,S2008-02,S1996-03,S2011-04,S1999-05}, holographic imaging~\cite{HI1997,AS2019,PL1980,L2007,CL2002}, augmented reality (AR)~\cite{JM2010,HK2014,AM2016}, chirp amplification~\cite{AB1997,AG1998,NF2005,HT2003,YS2017}, etc. However, based on such etching processes, blazed and holographic gratings, for instance, have a set grating constant, implying the challenge of dynamic manipulation of optical properties.

Electromagnetically induced grating (EIG)~\cite{H1998, M1999, HC2019, G2022}, based on quantum coherence techniques such as electromagnetically induced transparency (EIT)~\cite{H1990, H1997, O2024}, was proposed by Xiao's Group in 1998. The optical properties of EIG (e.g., grating constants) are dynamically tunable by standing wave modulation techniques. Combined with cross-phase modulation, high $\pm $ 1st-order diffraction efficiencies can be attained with low losses in the following EIG schemes~\cite{ZFX2013, DY2017, YY2019, CS2011, VA2015, GHJ2018, MDZ2021}. In recent years, the combination of non-Hermitian optical modulation techniques, including optical Parity-time symmetric ($\mathcal{PT}$)~\cite{REG2007, B2007, GL2013, MDD2019, YJP2020} and asymmetric ($\mathcal{APT}$) modulation~\cite{JHW2014, PP2016, YF2017}, can achieve dynamically tunable asymmetric diffraction in a symmetric structure. Some schemes have combined EIGs with Rydberg interactions to leverage non-local nonlinearity~\cite{SA2016, LYM2016, BF2018, MD2019}. These shemes consider the impact of photon statistical properties on the diffraction characteristics of gratings. These works greatly enrich the spectral principles and theoretical reserves of dynamically tunable gratings.

However, photons carry various degrees of freedom, including orbital angular momentum (OAM)~\cite{BS2003, AS2022, JJL2022, TZG2022, WA2023}, spin angular momentum, squeezed states, and polarization states, except for frequency, amplitude, and phase information. Some works combined optical diffraction with polarization states have been reported in waveguides and liquid crystals ~\cite{ZH1998, LY2004, A2007, D2017}. Research has also been conducted on the influence of polarization states on the diffraction characteristics of optical fields, employing the optical metasurface technique~\cite{PA2015, AA2015, MJP2017, RN2019, YR2023}. However, exploration of polarization state manipulation within EIG structures (high-performance dynamic control platforms) remains relatively limited~\cite{EPIG1, EPIG2, EPIG3}. It is worth noting that the discussion of polarization states here differs from that of ordinary and extraordinary beams discussed in optical super-surfaces or birefringent crystals.

Inspired by this, we present a theoretical scheme of EIG with diffraction properties dependent on the polarization state of the optical field. Implemented in an ultra-cold $^{87}Rb$ atomic ensemble, this grating incorporates both non-local and non-Hermitian optical modulation techniques. The non-Hermitian optical modulation can effectively control grating's symmetric and asymmetric optical diffraction and beam splitting. We introduce the Rydberg state into our model to achieve non-local optical modulation. Benefit from the Rydberg blockade effect, we can switch the optical response of the atomic system, utilizing the dynamic modulation of optical pulse trains.  
Furthermore, channels are established for probe photons with different polarization states (left- and right-handed circularly polarized), applying Zeeman sublevels (hyperfine structure). These two channels associated with incident polarization states can be individually controlled by dynamic light pulses and a variety of \textit{polarization-dependent} diffraction modes generate in this scheme. Our work provides a novel idea to utilize the polarization degree of freedom within EIG structures and gives insights for the control of versatile non-Hermitian optical devices.

This paper is organized through the following Sec.~\ref{PartII}, where we describe the basic model and theoretical scheme for dynamically switching the optical response of atomic system by Rydberg blockade effect. Two types of spatial modulation methods are also discussed here. We further consider selecting an appropriate level structure to achieve \textit{polarization-dependent} optical response channels. In Sec.~\ref{PartIII}, we discuss the multiple polarization-depndent diffraction modes and a non-local control for various diffraction modes by controlled Rydberg excitation. We summarize, at last, our conclusions in Sec.~\ref{PartIV}.

\begin{figure}[ptb]
\includegraphics[width=0.48\textwidth]{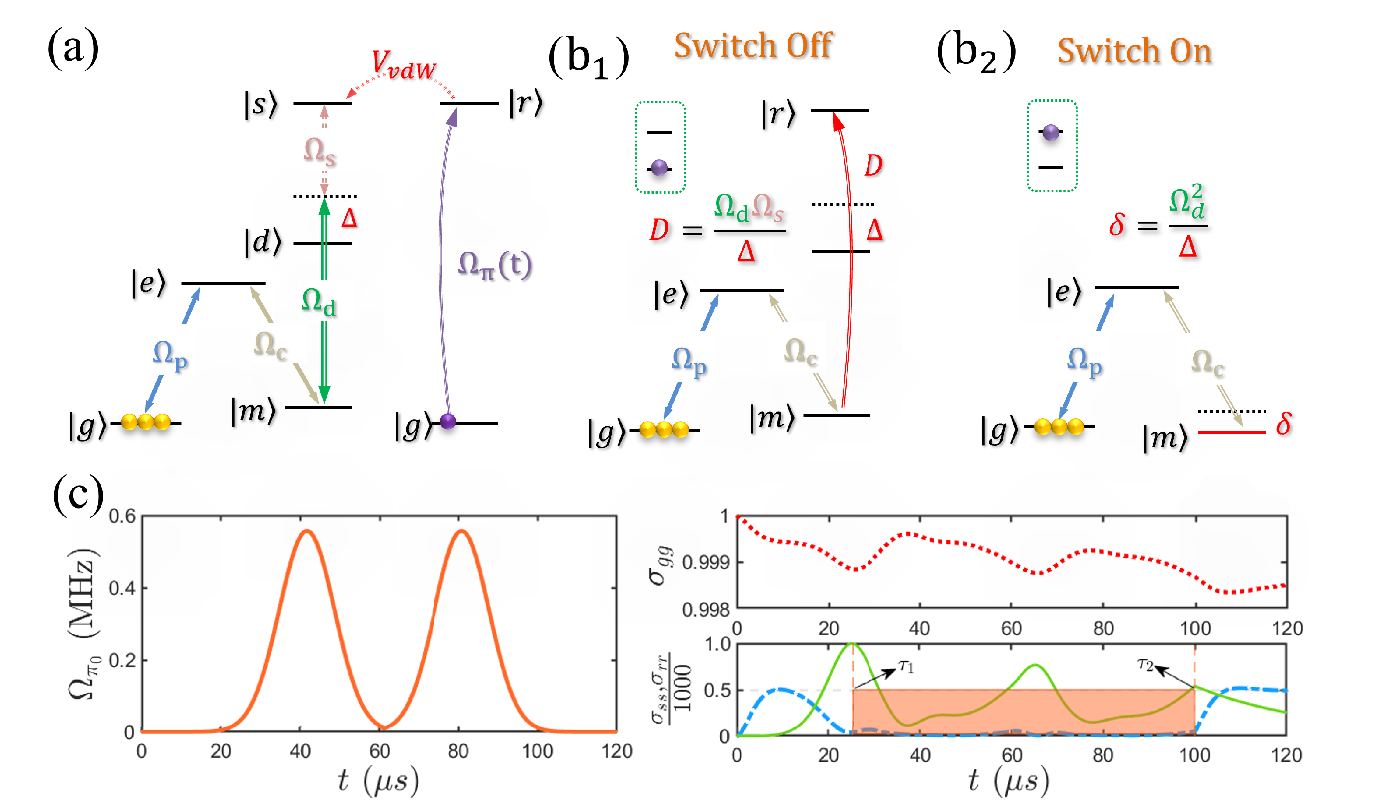}
\caption{Basic atomic energy level structure (a) and schematic diagram (equivalent energy level structures) for two stages (b$_1$) and (b$_2$). Continuously applying two gate pulses (c), the dynamic optical response of the system switches from that of equivalent $\mathcal{N}$-type four level structure (b$_1$) to that of $\Lambda$-type three-level system (b$_2$). Within the time region $\tau_1<t<\tau_2$  (red area), the population of state $|s\rangle$ ($\sigma_{ss}$, blue dashed curve depicted in the right panel of (c)) tends to zero, corresponding to ``Switch on'' stage (b$_2$) and decoupling of state $|s\rangle$. On the contrary, without the injection of gate pulses, the system remains in a steady state for most of the time, corresponding to ``Switch off'' stage (b$_1$).}
\label{Fig1}
\end{figure}

\section{The Basic Model}\label{PartII}

We aim to exploit non-Hermitian optical modulation to develop a \textit{polarization-dependent} scheme that can effectively control symmetric and asymmetric optical diffraction.~Before that, we first focus on dynamic modulation to switch the optical response of the atomic system, utilizing the pulse-triggered Rydberg interactions illustrated in Fig.~\ref{Fig1}.

First, we consider ultra-cold atomic gas driven by three classical laser fields (slow-varying amplitudes $\mathcal{E}_{c}$, $\mathcal{E}_{d}$ and $\mathcal{E}_{s}$). Travalling wave (TW) field $\omega_{c}$ ($\omega_{s}$) with Rabi frequency $\Omega_{c} = \mathcal{E}_{c}\cdot\wp_{me}/2\hbar$ ($\Omega_{s} = \mathcal{E}_{s}\cdot\wp_{ds}/2\hbar$) acts upon transition $|e\rangle\leftrightarrow|m\rangle$ ($|d\rangle\leftrightarrow|s\rangle$, $|s\rangle$ is a Rydberg state).~Standing wave (SW) field ($\omega_d$) couples the state $|m\rangle$ to intermediate excited state $|d\rangle$ non-resonantly with Rabi frequency $\Omega_{d}= \mathcal{E}_{d}\cdot\wp_{dm}/2\hbar$.~Weak field $\hat{\Omega}_p(z)=\hat{\mathcal{E}}_{p}(z) \cdot\wp_{ge}\sqrt{\omega_p/(2\hbar\varepsilon V)}$ with frequency ($\omega_p$) probe transition $|g\rangle\leftrightarrow|e\rangle$, where $\hat{\mathcal{E}}_{p}(z)$ and $V$ are the local probe amplitude operator and probe quantum volume.~Here, $\omega_{\mu\nu}$ and $\wp_{\mu\nu}$ describe the transition frequencies and the transition dipole moments ($\mu\in\{g, m, d, s\}$ and $\nu\in\{e, m\}$).~The detuning of probe field is defined as $\delta_{p}=\omega_p-\omega_{ge}$ with other corresponding detunings as $\delta_{c}=\omega_{c}-\omega_{me}$, $\Delta=\omega_{d}-\omega_{md}$, and $\delta_s=\omega_{s}-\omega_{ds}$.~The gate Rydberg state $|r\rangle$ can be excited by gate pulse ($\Omega_{\pi}(t)$, Ramman process) from ground state $|g\rangle$.

Under the far detuning condition $\Delta\gg\Omega_{d}$ and two photon resonance $\delta_{ms}=\Delta+\delta_s\simeq 0$, the system can be reduced to a quasi-$N$-type configuration with effcetive coupling $D=\Omega_d\cdot\Omega_s/\Delta$, illustrated in Fig.~\ref{Fig1}(b$_1$).~With the rotating-wave and electric-dipole approximations, the system can be naturally described by an effective interaction Hamiltonian $\mathcal{H}_{eff}=\mathcal{H}^{qN}_I+\mathcal{V}_{vdW}$, including an interaction Hamiltonian $\mathcal{H}^{qN}_I$, as well as van der Waals type dipole-dipole interactions between atoms $\mathcal{V}_{vdW}$:
\begin{align}\label{H_eff}
\mathcal{H}^{qN}_I & =  \hbar\sum_{k=1}^N\begin{bmatrix} 
0               & \Omega_p^{\dag} & 0              & 0              & \Omega_{\pi}^{*}(t)\\
\Omega_p        & -\delta_p       & \Omega_c^{*}   & 0              & 0 \\
0               & \Omega_c        & -\delta_{gm}   & D^{*}(x)       & 0 \\
0               & 0               & D(x)           & -\delta_{gs}   & 0 \\
\Omega_{\pi}(t) & 0               & 0              & 0              & -\delta_r
\end{bmatrix},\\
\mathcal{V}_{vdW} & = \hbar \sum_{i,j}^N\frac{C_6(s,r)}{R^6_{ij}}\sigma^{i}_{ss}\sigma^{j}_{rr},
\end{align}
where $\sigma^{i}_{ss}$ ($\sigma^{j}_{rr}$) is population of Rydberg state $|s\rangle_{i}$ ($|r\rangle_{j}$) for the $i$-th ($j$-th) atom in the ensemble. Here $C_6(s, r)$ is $vdW$-type interaction coefficient between state $|s\rangle$ and $|r\rangle$, with $\emph{R}\equiv(\emph{r}_i-\emph{r}_j)$ being the relative position vector between an atom at the position $\emph{r}_i$ and $\emph{r}_j$ [See panel (a) in Fig.~\ref{Fig1}].~Actually, we ignore the self-interaction term ($U_{self}\propto \sum_{i,j}^N\frac{C_6(s,s)}{R^6_{ij}}\sigma^{i}_{ss}\sigma^{j}_{ss}$) by choosing suitable Rydberg states ($C_6(s,s)\ll C_6(s,r)$) and providing detuning complement $\delta_r$ scaling of GHz. The following the Heisenberg-Langevin equations govern the dynamic evolution of the system:
\begin{align}\label{Eq_HL}
 \partial_t\hat{\sigma}(z, t) & =\frac{i}{\hbar}[\mathcal{H}_{eff}, \hat{\sigma}]+\mathcal{L}[\hat{\sigma}(z, t)],
 \end{align}
where $\mathcal{L}[\hat{\sigma}]=\sum_jc_j\hat{\sigma}c_j^{\dag}-\frac{1}{2}(c_j^{\dag}c_j\hat{\sigma}+\hat{\sigma}c_j^{\dag}c_j)$ with $c_j=\sqrt{\Gamma_j}\hat{\sigma}$ (e.g., $c_{eg}=\sqrt{\Gamma_{eg}}\hat{\sigma}_{eg}$). Before the gate pulse operation, $\Omega_{\pi}(t)=0$, the susceptibility is obtained as $\chi^{qN}(\omega, z)=\frac{2g^2}{\omega_p}\alpha^{qN}(\omega,z)\rho(z)$ with
\begin{align}\label{Eq_chi_alpha}
  \alpha^{qN} (\omega,z)& =\frac{i\gamma_{e}[\gamma_{m}^{\prime}\gamma_{s}^{\prime}+D^{*}D]}{\gamma_{e}\gamma_{m}^{\prime}\gamma_{s}^{\prime}+\gamma_{m}^{\prime}\Omega_{c}^{*}\Omega_{c}+\gamma_{e}D^{*}D}.
 \end{align}
 Here $\gamma_{e}^{\prime}=\gamma_e+i\delta_{p}$, $\gamma_{m}^{\prime}=\gamma_{m}+i(\delta_{p}-\Delta_{m})$, and $\gamma_{s}^{\prime}=\gamma_{s}+i(\delta_{p}-\Delta_{m}+\delta_{s})$ are also introduced to denote the complex dephasing rates for convenience, with $\gamma_{\mu\nu}=\sum_{k}(\Gamma_{\mu k}+\Gamma_{\nu k})/2$.~The small $vdW$-induced average frequency shift $\langle\hat{s}(z)\rangle$ (self interaction), which can be absorbed into $\delta_{s}= \delta_{s_0}-\langle\hat{s}(z)\rangle$ suitably. 
 
 When the gate pulse is incident, due to $\langle\hat{s}(z)\rangle\to\infty$, the Rydberg excitation of state $|r\rangle$ will blockade state $|s\rangle$, resulting in the dynamic decoupling of state $|s\rangle$ from the system [See Fig.~\ref{Fig1}(b$_2$)]. In this scenario, within the blockade region of state $|r\rangle$ (blockade 
 radius $R_b(r,s)\simeq\sqrt[6]{\frac{C_6(r,s)}{w}}$), the atomic level structure degenerates into $\Lambda$-type three-level configuration with a large frequency detuning (ac Stark shift $\delta=\Omega_d^2/\Delta$). It corresponds to the effective Hamiltonian
 \begin{align}\label{H_eff_2}
  \mathcal{H}^{3l}_I & =  \hbar\sum_{k=1}^N\begin{bmatrix} 
  0               & \Omega_p^{\dag} & 0              \\
  \Omega_p        & -\delta_p       & \Omega_c^{*}   \\
  0               & \Omega_c        & -\delta_{gm}+\delta   \\
  \end{bmatrix},
\end{align}
with the susceptibility $\chi(\omega,z)  =\frac{\rho(z)\wp_{ge}}{\gamma\hbar\epsilon_0}\alpha^{3\Lambda}(\omega,z)$ ($ \alpha^{3\Lambda}=\gamma_e\sigma_{ge}/\Omega_{p}$) and
\begin{align}\label{rho_1}
 \alpha^{3\Lambda} (\omega,z) & \simeq \frac{i\gamma_e}{\gamma_e+i[\delta_{p}+\Omega_d^2/\Delta_d]+\Omega_{c}^2/\gamma_{m}^{\prime}}
\end{align}
obtained in the stationary regime.~This dynamical decoupling process is well demonstrated by Fig.~\ref{Fig1}(c). Within the time interval $\tau_1 < t < \tau_2$, the population of state  $|s\rangle$ ($\sigma_{ss}$) rapidly decreases, thereby decoupling from the system.

\subsection{Diffraction}\label{Dif}

Aiming to produce diffraction patterns in the $x$ direction, we spatially modulate the ensemble with the coupling field as SW $\Omega_d(x)=\Omega_d/\sqrt{2}\cos[2\pi\lambda_d(x-x_0)/a+\psi]$ along $x$ direction (with detuning $\omega_{dm}-\omega_d=\Delta$ and small angle $\theta$). The dressing field also includes TW component ($\Omega_d/\sqrt{2}$ and $\omega_{dm}-\omega_d=-\Delta$) along $z$ direction. Meanwhile, there is a phase shift of $\pi/4$ between the nodes of the SW field and $z$ axis (see Fig.~\ref{Fig2}(c) and follow the method in \cite{JHW2014, LoopEIG}). Here $\lambda_{d}$ ($a$) is the wavelength (spatial period) of the coupling field with $\psi$ being controllable phase of detuning. Before the gate pulse operation, the atoms with a quasi-$N$ type four level structure experience a SW field amplitude modulation in the form of $D(x)=\Omega_s\Omega_{d_0}\cos[2\pi\lambda_d(x-x_0)/a]/\Delta$. Thus, the real (imaginary) part of probe susceptibilities $\chi^{\prime}(x,z)$ ($\chi^{\prime\prime}(x,z)$) is used to describe the dispersion (absorption/gain) property. The transmission function of a probe beam at $z=\mathcal{L}$ takes the form
\begin{align}
T(x)=A(x)\cdot P(x),
\label{Eq_T}
\end{align}
where $A(x)$ = $e^{-\int_{0}^{\mathcal{L}}k_p\chi^{\prime\prime}(x,z)dz}$ ($P(x)$ = $e^{i\int_{0}^{\mathcal{L}}k_p\chi^{\prime}(x,z)dz}$) denotes the amplitude (phase) component with $k_p=2\pi/\lambda_{p}$ ($\lambda_{p}$) being the probe wave vector (wavelength). The Fourier transformation of $T(x)$ then yields the Fraunhofer or far-field intensity diffraction equation
\begin{align}\label{Eq_Ip}
\mathcal{E}^{\mathcal{L}}(\theta_n) & =\int_{-a/2}^{+a/2}{dx [T(x)e^{-i2{\pi}xR\sin{(\theta_n})}]},\\
I(\theta_n) & =\frac{{\vert}\mathcal{E}^{\mathcal{L}}(\theta_n){\vert}^{2}\sin^{2}[M{\pi}R\sin(\theta_n)]}{M^{2}\sin^{2}[{\pi}R\sin(\theta_n)]},
\end{align}
with $R=a/{\lambda_{p}}$. In addition, $\theta_n$ denotes the $n$th order diffraction angle of probe photons to the $z$ direction while $M$ represents the ratio between the beam width $\varpi_B$ and the grating period $a$ ($M=\varpi_B/a$). The $n$th-order diffracted probe field will be found at an angle determined by $n=R\sin{\theta_n}\in{(\ldots,-1,0,+1,\ldots)}$. At this point, a normal far-field diffraction angle spectrum and diffraction pattern of EIG can be attained.

Conversely, after the gate pulse operation, atoms will decoupled from the Rydberg state $|s\rangle $ would exhibit spatial modulation relative to the detuning $\delta(x) = \delta_0 \cdot \sin[4\pi \lambda_d (x - x_0)/a + \psi]$, following the approach outlined in Ref.~\cite{JHW2014, LoopEIG}. Here $\delta_0=\Omega_d^{*}\Omega_d/\Delta$ is ac Stark shift. In this case of detuning spatial modulation $\delta_{0}\neq 0$, we can expend the real amd imaginary parts of susceptibility as polynomials $\chi^{\prime}(x)=\sum_{k=0}^{n^{\prime}_{\zeta}}  C_{n^{\prime}}  \delta^{n^{\prime}}(x)$ and $\chi^{\prime\prime}(x)=\sum_{k=0}^{n^{\prime\prime}}  C_{n^{\prime\prime}} \delta^{n^{\prime\prime}}(x)$ for $\delta(x)$, where $C_{n^{\prime},n^{\prime\prime}}$ are polynomial coefficients ($n, n^{\prime}, n^{\prime\prime}\in\mathbb{Z}$). We can determine whether the optical medium satisfies $\mathcal{PT}$-antisymmetry by judging the parity of the real and imaginary parts with respect to the spatial position $x$ based on these polynomials (See \eqref{A1} in Appendix). So, in this scenario, if we discuss the diffraction of the probe field, we can obtain an asymmetric diffraction spectrum characteristic of a non-Hermitian grating.

\subsection{\textit{Polarization-dependent} Optical Response}\label{POR}
\begin{figure}[ptb]
  \includegraphics[width=0.48\textwidth]{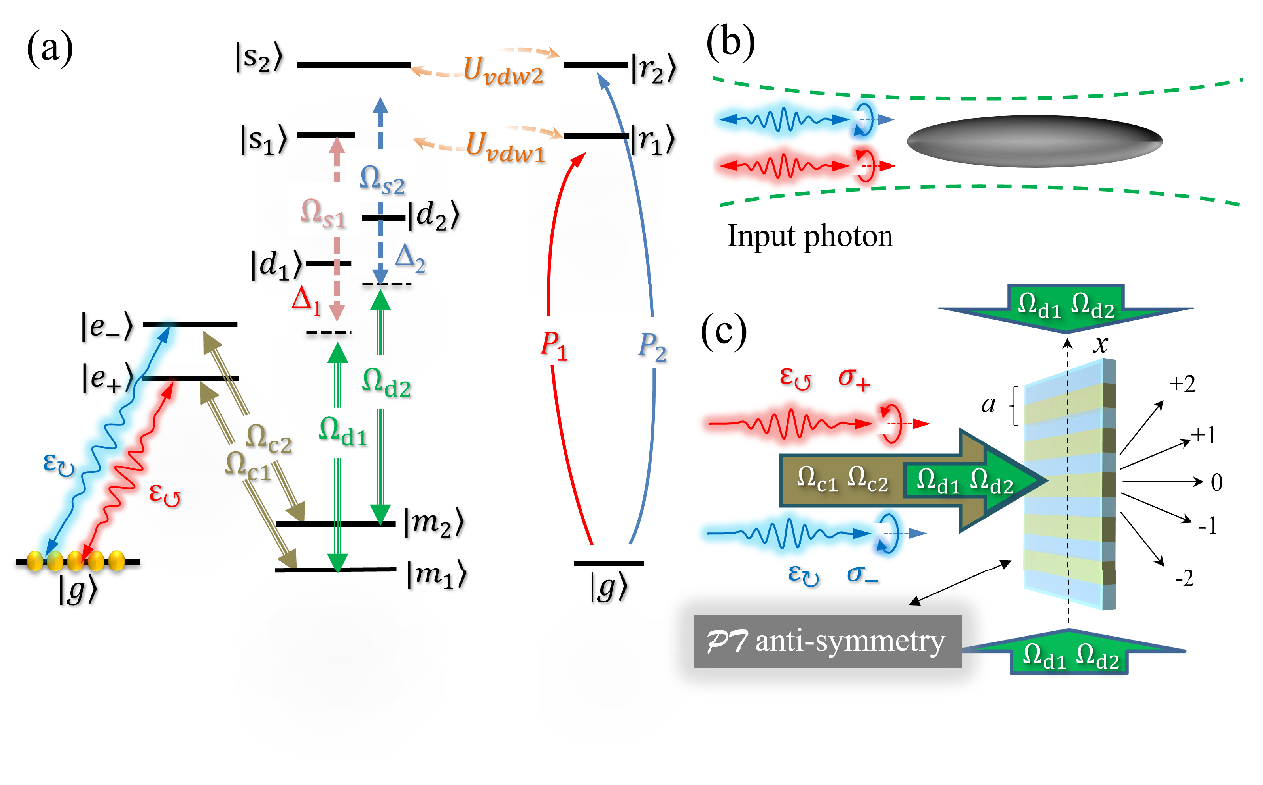}
  \caption{Atomic energy level structure for left/right-cirlar polarization channels (a) and schematic diagram of polarization selective non-Hermitian atomic grating (b, c). The spatial modulation scheme ($\mathcal{PT}$ antisymmetry) follows the method in references \cite{JHW2014, LoopEIG} and is also discussed in Sec.~\ref{Dif}.}
  \label{Fig2}
  \end{figure}
Thirdly, we consider selecting an appropriate level structure to achieve optical response channels depending on verious polarization states (left- and right-circular polarized light, as well as linearly polarized light) of incident beams. We utilize the Zeeman sublevels to construct two symmetrically coupled channels, as shown in Fig.~\ref{Fig2}(a), $|g\rangle \to |e_{\pm}\rangle \to |m_{\xi}\rangle \to |d_{\xi}\rangle \to |s_{\xi}\rangle$, where $\xi=\{1,2\}$ ($|e_{+}\rangle$ and $|e_{-}\rangle$) correspond left- and right-circular polarization channels ($\circlearrowleft$ and $\circlearrowright$), respectively. Weak fields $\hat{\Omega}_{\circlearrowleft,\circlearrowright}(z)=\hat{\mathcal{E}}_{p}(z) \cdot\wp_{ge_{\pm}}\sqrt{\omega_p/(2\hbar\varepsilon V)}$ with different circular polarization ($\sigma_{+}$ and $\sigma_{-}$) but the same frequency ($\omega_p$) probe transitions $|g\rangle\leftrightarrow|e_{\pm}\rangle$.~The classical control fields $\omega_{c_1,c_2}$ ($\Omega_{c_1,c_2} = \mathcal{E}_{c_1,c_2}\cdot\wp_{me_{\pm}}/2\hbar$) act upon transitions $|m_{1,2}\rangle\leftrightarrow|e_{\pm}\rangle$.~Additionally, two driving fields that are spatially modulated couple non-resonantly the states $|m_{\xi}\rangle$ to intermediate excited states $|d_{\xi}\rangle$ with Rabi frequency $\Omega_{d_{\xi}}= \mathcal{E}_{d_{\xi}}\cdot\wp_{dm_{\xi}}/2\hbar$. Here, $\omega_{\mu\nu}$ and $\wp_{\mu\nu}$ describe the transition frequencies and the transition dipole moments ($\mu\in\{g, m_{\xi}, d_{\xi}\}$ and $\nu\in\{e_{+},e_{-}\}$). The detuning of the left (right) circular polarization probe field is defined as $\delta_{\circlearrowleft}=\omega_p-\omega_{ge_{+}}$ ( $\delta_{\circlearrowright}=\omega_p-\omega_{ge_{-}}$) with other corresponding detunings as $\delta_{c_{\xi}}=\omega_{c_{\xi}}-\omega_{m_{\xi}e_{\pm}}$ and $\Delta_{\xi}=\omega_{d_{\xi}}-\omega_{m_{\xi}d_{\xi}}$. 

Similarly, without the gate pulse operation, the susceptibilities for different polarized photons are obtained as
 $\chi_{\zeta}^{qN}(\omega, z)=\frac{\wp_{ge}}{\gamma\hbar\epsilon_0}\alpha^{qN}_{\zeta}(\omega,z)\rho(z)$ with
\begin{align}\label{Eq_chi_alpha}
 \alpha_{lc}^{qN} (\delta_{\circlearrowleft},z)& =\frac{i\gamma_{\circlearrowleft}[\gamma_{m_1}^{\prime}\gamma_{s_1}^{\prime}+D^{*}_{1}D_{1}]}{\gamma_{\circlearrowleft}\gamma_{m_1}^{\prime}\gamma_{s_1}^{\prime}+\gamma_{m_1}^{\prime}\Omega_{c_{1}}^{*}\Omega_{c_{1}}+\gamma_{\circlearrowleft}D^{*}_{1}D_{1}},\nonumber\\
 \alpha_{rc}^{qN} (\delta_{\circlearrowright},z)& = \frac{i\gamma_{\circlearrowright}[\gamma_{m_2}^{\prime}\gamma_{s_2}^{\prime}+D^{*}_{2}D_{2}]}{\gamma_{\circlearrowright}\gamma_{m_2}^{\prime}\gamma_{s_2}^{\prime}+\gamma_{m_2}^{\prime}\Omega_{c_{2}}^{*}\Omega_{c_{2}}+\gamma_{\circlearrowleft}D^{*}_{2}D_{2}},
\end{align}
where $\gamma_{s_{\xi}}^{\prime}=\gamma_{s_{\xi}}+i[\delta_{\circlearrowleft,\circlearrowright}-\delta_{c_{\xi}}+\delta_{s_{\xi}}]$ ($\zeta\in\{lc, rc\}$). And different dark states can be attained under the zero detuning condition ($\delta_{s_{\xi}}=\delta_{c_{\xi}}=0$), $|$DS$\rangle_{\circlearrowleft}=\frac{\Omega_{c_1} D_{1}}{N_{lc}}|g\rangle -\frac{\Omega_{\circlearrowleft} D_{1}}{N_{lc}}|m_1\rangle+\frac{\Omega_{\circlearrowleft} \Omega_{c_1}}{N_{lc}}|s_1\rangle$ and $|$DS$\rangle_{\circlearrowright}=\frac{\Omega_{c_2} D_{2}}{N_{rc}}|g\rangle -\frac{\Omega_{\circlearrowright} D_{1}}{N_{rc}}|m_2\rangle+\frac{\Omega_{\circlearrowright} \Omega_{c_2}}{N_{rc}}|s_2\rangle$, with $N_{lc,rc}=\sqrt{D^2_{\xi}(\Omega^2_{\circlearrowright,\circlearrowleft}+\Omega_{c_{\xi}}^2)}$.

With control gate pulses coupling to states $|r_1\rangle$ and $|r_2\rangle$ (Rydberg states), we can obtain two individual controllable channels for different circle polarization beams. The selection of Rydberg states, characterized by $C_6(r_{\xi}, s_{\xi}) \gg C_6(s_{\xi}, s_{\xi})$ and the depreciation of cross-interactions, renders the Rydberg excitation of gate states $|r_{\xi}\rangle$. Actually, when the blockade effect occurs, the steady-state optical response in the corresponding channel corresponds to $\Lambda$-type suspectibilities 
\begin{align}\label{rho_1}
  \alpha_{lc}^{3\Lambda} (\delta_{\circlearrowleft},z) & =\frac{i\gamma\gamma_{m_1}}{\gamma_{\circlearrowleft}\gamma_{m_1}+\Omega_{c_{1}}^2},\nonumber\\
  \alpha_{rc}^{3\Lambda} (\delta_{\circlearrowright},z) & = \frac{i\gamma\gamma_{m_2}}{\gamma_{\circlearrowright}\gamma_{m_2}+\Omega_{c_{2}}^2}.
 \end{align}
 Naturally, continuing to discuss the diffraction of beams with different polarization states can be analogous to Eqs.~(7-9), using polarization-state-related physical quantities ($T_{\zeta}$, $A_{\zeta}$, $P_{\zeta}$, $\mathcal{E}_{\zeta}^{\mathcal{L}}, I_{\zeta}$) for description. The corresponding spatial modulation are then expressed as
 \begin{align}
  \delta_{l}(x) = \delta_{l0} \cdot \sin[4\pi \lambda_d (x - x_0)/a + \psi_1],\nonumber\\
  \delta_{r}(x) = \delta_{r0} \cdot \sin[4\pi \lambda_d (x - x_0)/a + \psi_2],
  \label{Eq_delta_c}
  \end{align}
where $\psi_{\xi}$ are controllable phases with ac Stark shift $\delta_{l0,r0}=\Omega_{d_{\xi}}^{*}\Omega_{d_{\xi}}/\Delta_{\xi}$ for different polarization channels ($\xi\in\{1,2\}$).
 
\section{Results and discussion}\label{PartIII}

This section discusses the grating's far-field diffraction (Fraunhofer diffraction or under paraxial approximation) and its manipulation, employing real atomic levels and parameters that closely emulate experimental conditions. We consider an ensemble of ultra-cold $^{87}$Rb atoms in an elongated trap of length $\mathcal{L}\simeq 200\mu$m. The ground $|g\rangle$, metastable states $|m_{1,2}\rangle$ and excited states $|e_{\pm}\rangle$ of the medium atoms would correspond to suitable sub levels of the $5S_{1/2}|F=2, m_F=0,-2,+2\rangle$ and $5P_{1/2}|F=1,m_F=-1,+1\rangle$.~For convenience, the selection of Rydberg states will be discussed later.~The quantization direction is taken along the $z$ axis. We choose the atomic density $N=1.0\times 10^{11}$ cm$^{-3}$ with the probe wavelength $\lambda_p=795$ nm and transition dipole moment $\wp_{ge_{\pm}}=2.534\times 10^{-29}$ C$\cdot$m.
 
\subsection{Diffraction for \textit{Polarization-dependent} $\Lambda$-type channels}\label{Sect.II}
\begin{figure}[ptb]
\includegraphics[width=0.48\textwidth]{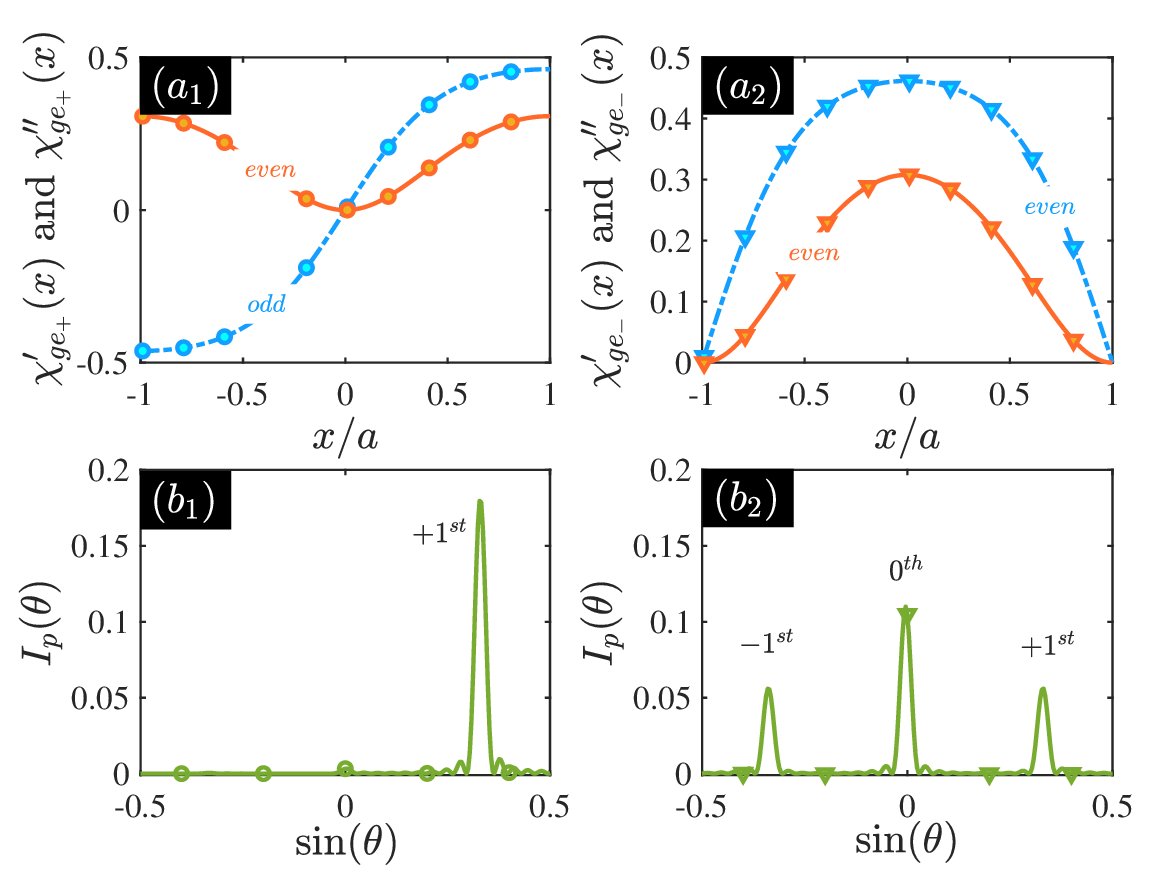}
\caption{Polarization dependent diffraction \textit{Mode I}: Absorption (orange-solid) and dispersion (blue-dashed) curves versus $x$ of left-circlar (circle-marked) and right-circlar (triangle-marked) polarization probe beams in (a$_1$) and (a$_2$). 
Panel (b$_1$) and (b$_2$) show diffraction intensity $I_p(\theta)$ vesurs diffraction angle $\theta$ for $\circlearrowleft$ and $\circlearrowright$ probe fields, with $\Omega_{c_{\xi}}=3.0\times 2\pi$ MHz, $\Omega_{d_{\xi}}=2.0\times 2\pi$ MHz, $\Delta_{\xi}=20\times 2\pi$ MHz ($\xi=1, 2$), $\psi_1=0$, and $\psi_2=\pi/2$.}
\label{Fig3}
\end{figure}

We starting from discussing the basic diffraction properties of polarization gratings without considering Rydberg states (assuming Rydberg states decoupled from the system).
Despite having equal frequencies ($\omega_{\xi}\simeq\omega_p$), the varied polarization probe photons interact on the distinct effective level structures, resulting in diverse linear responses. It indicates a variety of polarization state-dependent behaviors of atomic ensembles, including absorption, dispersion, and scattering, which enables the implementation of a controlled \textit{polarization-dependent} electromagnetically induced optical diffraction scheme.  Three diverse diffraction modes are depicted, depending on the polarization of the input beams:

\textit{Mode I}. Asymmetric diffraction for the left circle ($\circlearrowleft$, $lc$ for simplify) beam but symmetric diffraction for the right circle ($\circlearrowright$, $rc$ for simplify) beam are illustrated in Figure~\ref{Fig3}. As mentioned above, setting the phase to $\psi_1=n\pi$ engenders a distinctive interaction for left-circularly polarized photons with the ensemble atoms, wherein the effective susceptibility manifests a spatially \textit{odd}-function real part $\chi_{lc}^{\prime}(-x)=-\chi_{lc}^{\prime}(x)$ alongside a spatially \textit{even}-function imaginary part $\chi_{lc}^{\prime\prime}(-x)=\chi_{lc}^{\prime\prime}(x)$, conforming to PT-antisymmetry ($\mathcal{APT}$), as illustrated in Fig.~\ref{Fig3}(a$_1$). Consequently, this leads to perfect asymmetric diffraction (only diffracted asymmetrically into positive angles $\theta>0$) upon passing through the atomic ensemble, as illustrated in Fig.~\ref{Fig3}(b$_1$). Conversely, the susceptibility for $rc$ photons $\chi_{rc}(-x)=\chi_{rc}(\delta\cos[\pi x/a])=\chi_{rc}(x)$ is \textit{even}-function of axial $x$ [see Fig.~\ref{Fig3}(a$_2$)], which corresponds to symmetric diffraction in Fig.~\ref{Fig3}(b$_2$), with $\psi_{2}=\pi/2$.

\begin{figure}[ptb]
\includegraphics[width=0.48\textwidth]{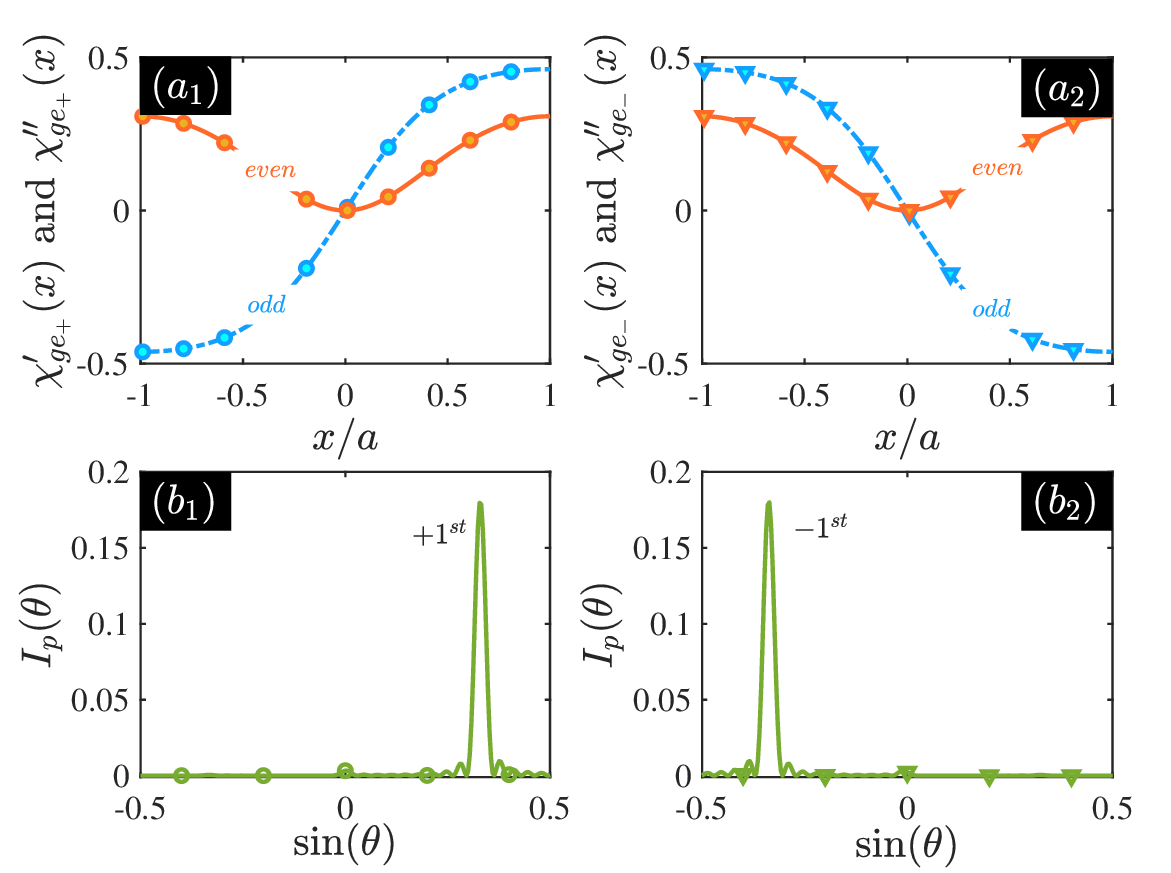}
\caption{Polarization dependent diffraction \textit{Mode II}: Absorption (orange-solid) and dispersion (blue-dashed) curves versus $x$ of left-circlar (circle-marked) and right-circlar (triangle-marked) polarization probe beams in (a$_1$) and (a$_2$). Panel (b$_1$) and (b$_2$) show diffraction intensity $I_p(\theta)$ vesurs diffraction angle $\theta$ for $\circlearrowleft$ and $\circlearrowright$ probe fields, with $\psi_1=0$, and $\psi_2=\pi$. Other parameters are the same as Fig.~\ref{Fig3}.}
\label{Fig4}
\end{figure}

\textit{Mode II}. Asymmetric diffraction in the different directions for distinct polarization photons is depicted in Figure~\ref{Fig4}, with the spatial modulation as $\Delta_{1}/2\pi=\Delta_{2}/2\pi=50.0$ MHz, $\psi_{1}=2n\pi$, and $\psi_{2}=(2n+1)\pi$ ($n\in \mathbb{Z}$).~A \textit{polarization-dependent} beam splitter is implied here, which depicts that photons with different circlar polarization states will be diffracted into separate channels (at lopsided positive or negative angles), shown as Fig.~\ref{Fig4}(b$_1$) and (b$_2$). The intensity distribution of the two polarization-dependent diffraction channels and the diffraction direction are different under these circumstances. 

Additionally, it is easily attained the same response of the two different channels ($\circlearrowleft$ and $\circlearrowright$) by modifying the sign of detuning $\Delta_{2}$ in \textit{Mode II} ($\Delta_{2}/2\pi=-50$ MHz) or setting $\psi_2=\psi_1=k\pi$ ($k=2n$ and $k=2n+1$ for both lopsided positive or negative angle diffraction), which is \textit{Mode III}. the same direction asymmetric diffraction under $\mathcal{PT}$-symmetric spatial modulation. 

\textit{Mode I} and \textit{II} demonstrate the beam-splitting capability of the non-Hermitian grating according to its polarization properties demonstrating that through the strategic configuration of the level structure, we possess the capability to precisely modulate the diffraction characteristics of polarized photons, merely by adjusting the phase responsible for inducing ac Stark shifts.

\subsection{\textit{Polarization-dependent} response under Rydberg control}

Next, we will discuss the cases considering Rydberg states and interactions integrated into the system. Without considering Rydberg interactions, we can get quasi-four level structure ($\mathcal{N}$-type) optical response of the atomic ensemble including the effective coupling $D_{\xi}=-\Omega_{c_{\xi}}\Omega_{d_{\xi}}/\Delta_{\xi}$ separately for different polarization channels, depicted in Fig.~\ref{Fig2}. It is noteworthy that, since $\Omega_{d_{\xi}}(x)=\Omega_{d_{\xi}}\cos[2\pi x/a]$ functions versus $x$, the effective coupling within the new structure retains a periodic spatial modulation. Consequently, the system's spatial modulation transitions from detuning modulation to amplitude modulation, ensuring the emergence of another type of EIG structure.

Considering the corresponding detuning compensation $\delta_{s_{\xi}}$ (scale of GHz), an approximate two-photon resonance condition $\Delta_{\xi}+\delta_{s_{\xi}}-\langle \hat{s}_{s_{\xi}}\rangle \simeq 0$ is provided, after taking into account the relatively weak self-interaction of the $|s_{\xi}\rangle$ state, denoted as $U^{self}_{\xi}\propto C_6(s_{\xi}, s_{\xi})$. Choosing suitable Rydberg states ($C_6(r_{\xi}, s_{\xi}) \gg C_6(s_{\xi}, s_{\xi})$) renders the Rydberg excitation of gate states $|r_{\xi}\rangle$ ($\sigma_{r_{\xi}r_{\xi}}^{g}$) as the key point in controlling the ensemble optical response via van der Waals ($vdW$) interactions. This control is facilitated through $vdW$ interactions that operate across diverse channels, with the interaction strength $U_{vdW_{\xi}}\propto\frac{C_6(s_{\xi}, r_{\xi})}{|r_{t}-r_{g}|^6}\sigma_{r_{\xi}r_{\xi}}^{g}$. 

\begin{figure}[ptb]
\includegraphics[width=0.48\textwidth]{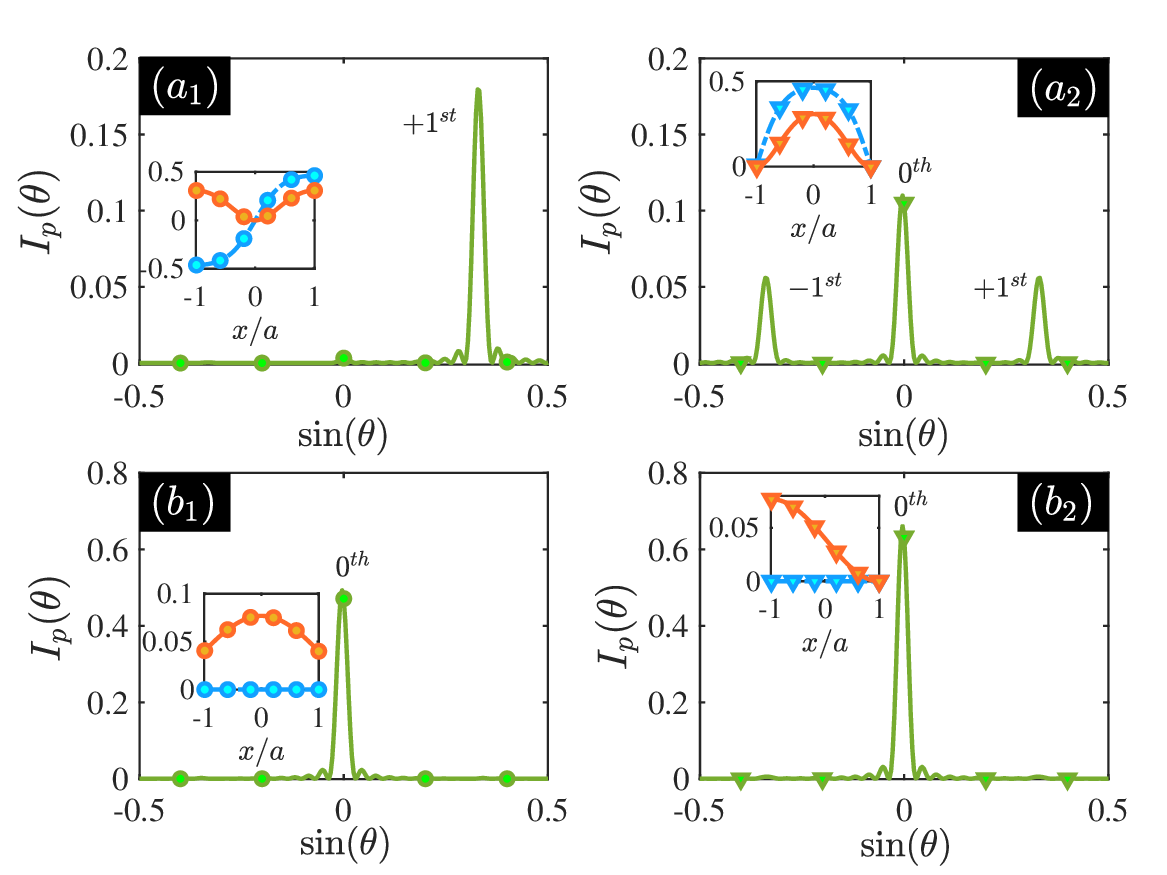}
\caption{Diffraction angle spectrum $I_p(\theta)$ for $\circlearrowleft$ and $\circlearrowright$ probe fields, showed in left/right two panels, respectively. Top two panels, (a$_1$) and (a$_2$), dispaly diffraction intensity $I_p(\theta)$ under the switch-on condition. The switch-off cases correspond to the bottom two panels, (b$_1$) and (b$_2$). The phase of detunings are set as $\psi_1=0$, and $\psi_2=\pi$. Other parameters are the same as Fig.~\ref{Fig3}.}
\label{Fig5}
\end{figure}

  \begin{figure}[ptb]
    \includegraphics[width=0.48\textwidth]{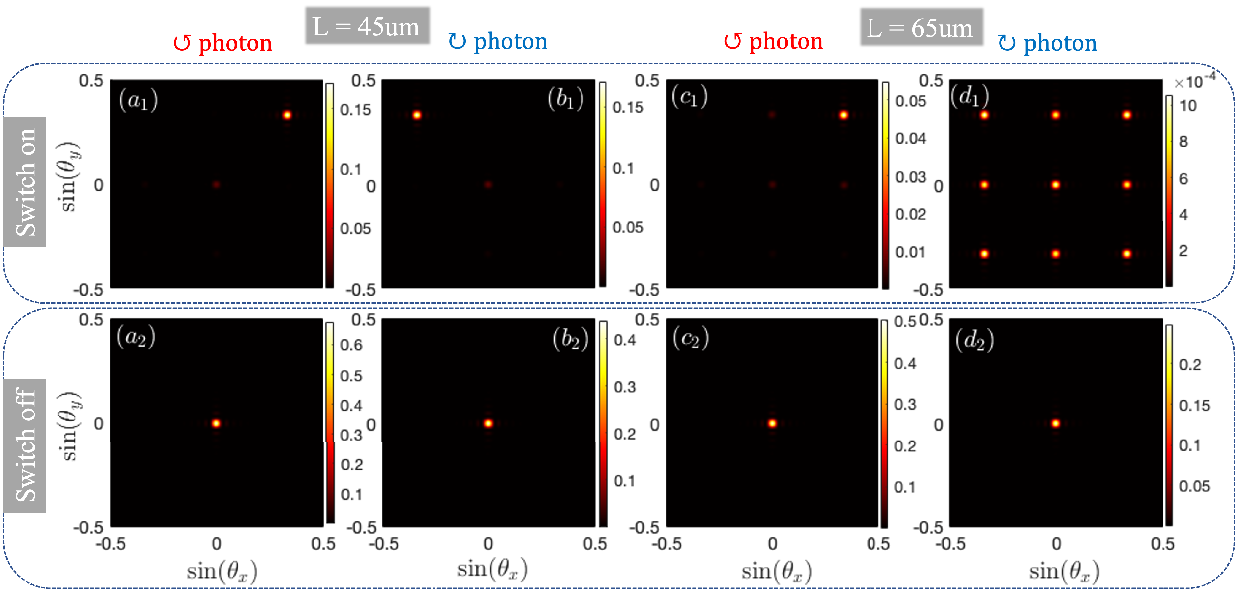}
    \caption{Two dimensional diffraction angle spectrum $I_p(\theta_x,\theta_y)$ for $\circlearrowleft$ and $\circlearrowright$ probe fields, under the different OD with $\mathcal{L}=45\mu$m (65$\mu$m) for the left (right) four panels. Top four panels, (a$_1$)- (d$_1$), dispaly diffraction intensity $I_p(\theta)$ under the switch-on condition. The switch-off cases correspond to the bottom four panels, (a$_2$)-(d$_2$). The phase of detunings are set as $\psi_1=0$, and $\psi_2=\pi$. Other parameters are the same as Fig.~\ref{Fig3}, except for $\Omega_{d_{\xi}}(x,y)=\Omega_{d_{0}}[\sin(\pi x/a+\psi_{x_{\xi}})+\sin(\pi y/a+\psi_{y_{\xi}})]$.}
    \label{Fig6}
    \end{figure}

\begin{table*}[htbp]
      \centering
      \caption{Rydberg atom states with the corresponding $vdW$ interaction coefficient $C_6$, lifetime $\tau$ (decay rate $\Gamma=\frac{1}{\tau}$) ignoring blackbody radiation (BBR), the Le Roy radius $R_{LR}$, two-type blockade radius ($R_b^{ss}$ self-interaction between state $|s_{\xi}\rangle$; $R_b^{rs}$ cross-interaction between state $|r_{\xi}\rangle$ and $|s_{\xi}\rangle$).}
      \begin{tabular}{ccccccccc}
      \hline\hline
       State & &$vdW$-type & interaction  &  &   Decay rate   &  LR radius & Blockade radius I.& Blockade radius II.\\
        $n$   & & $C_{6}(n,n^{\prime})$   & (GHz$\cdot\mu m^6$) &&  $\Gamma$ ($2\pi\cdot$kHz)   &   $R_{LR}$($\mu m$) & $R_{b}^{rr}$($\mu m$) & $R_{b}^{rs}$($\mu m$)\\
       \hline             & $n^{\prime}=40S_{1/2}$   & $44S_{1/2}$   & $41S_{1/2}$   & $45S_{1/2}$   &       &  & & \\
       \hline
      $40S_{1/2}$ & $-1$          & \textit{neg.} & $12$          & \textit{neg.} & 15.95 & 0.5 & 2.0 & 4.0\\
      $44S_{1/2}$ & \textit{neg.} & $-3$          & \textit{neg.} & $-16$          & 11.7  & 0.6 & 2.1 & 4.2\\
      \hline\hline
      \end{tabular}
        \label{tab1}
\end{table*}

In the absense of gate pulses (in form of Ramman process) to Rydberg spin ($\langle\hat{\sigma}_{gr_{\xi}}(\tau)\rangle= 0$), the blockade effect induced by $vdW$ interaction ($U_{vdW_{\xi}}$) does not occur. Fig.~\ref{Fig4}(b$_1$) and (b$_2$) illustrate the diffraction intensity distribution of photons with different polarizations utilizing distinct phases, $\psi_1=0$ and $\psi_1=\pi/2$, respectively. Photons are primarily diffracted into the $0^{th}$ order, which is reflected by the almost zero dispersive capability indicated by the blue curves in the subfigures.

Upon the excitation of the gate states $|r_{\xi}\rangle$, a blockade effect ensues for the corresponding state $|s_{\xi}\rangle$ of the ensemble atoms, with blockade radius $R_b(r_{\xi},s_{\xi})\simeq\sqrt[6]{\frac{C6(r_{\xi},s_{\xi})}{w}}$ under the laser line-width $w=3$MHz, leading to its decoupling from the optical field denoted by $\Omega_{s_{\xi}}$. Due to the large detuning condition $\Delta_{\xi}\gg\Omega_{d_{\xi}}$ (between $|m_{\xi}\rangle\leftrightarrow|d_{\xi}\rangle$), only with the small ac Stark frequency shift $\delta_{\xi}\simeq-\Omega_{d_{\xi}}^2/\Delta_{\xi}$ left. 
Hence, with the symmetric coupling condition $\Omega_{c_{1}}=\Omega_{c_{2}}=\Omega_{c}$ and $\Delta_{1}=\Delta_{2}=\Delta$, the identical medium linear susceptibilities for different polarization beams will be attained as Eqs.~\eqref{rho_1}, the consistent linear response. Including the diffraction properties after propagation through the atomic ensembles, it reverts to the scenario described in Sect.~\ref{Sect.II} [See Fig.~\ref{Fig5}(a$_1$) and (a$_2$)]. The collective excitation of the gate states ($|r_{\xi}\rangle$) is the switch for the system's \textit{Polarization-dependent} optical response.

In selecting Rydberg states, several guiding principles should be observed.~Foremost, it is important to limit the principal quantum number of the Rydberg states ($n_{s_{\xi}}$) to compensate for frequency shifts caused by self-interactions. 
Additionally, it is imperative that the interaction coefficient between the gate Rydberg state and the target state of the ensemble atoms significantly exceeds the self-interaction coefficients $C_6(n_{s_{\zeta}},n_{r_{\xi}})\gg C_6(n_{s_{\xi}},n_{s_{\xi}})$ ($\xi=1,2$), with the parameters $C_6(n_{s_1},n_{r_2})$ and $C_6(n_{s_2},n_{r_1})$ closely approximated to zero to prevent the introduction of crosstalk in the control process.~Accordingly, the states selected and their pertinent details are listed as $|s_1\rangle \equiv |40S_{1/2}, m_j = 1/2\rangle$, $|s_2\rangle \equiv |44S_{1/2}, m_j = 1/2\rangle$, $|r_1 \rangle\equiv |41S_{1/2}, m_j = 1/2\rangle$, and $|r_2\rangle \equiv |45S_{1/2}, m_j = 1/2\rangle$ with the related $vdW$ interaction coefficients and lifetimes detailed in Table \ref{tab1}, expect for the blockade radius $R_b(41S,40S)=4.0\mu m$ and $R_b(45S,44S)=4.2\mu m$.

A pair of Raman pulses, involved in the STIRAP (Stimulated Raman Adiabatic Passage) process, acts between the states $|g\rangle$ and $|s_{\xi}\rangle$. Due to the dipole blockade effect among $|s_{\xi}\rangle$ states, the medium forms an arrangement of blockade spheres with a blockade radius of 2.6 $\mu m$, within which at most a single collective excitation of the $|s_{\xi}\rangle$ state can exist. The interaction between $|s_{\xi}\rangle$ and $|r_{\xi}\rangle$ states results in the medium being segmented into a stack of blockade spheres to the $|r_{\xi}\rangle$ state, each having a blockade radius of approximately 4.0 $\mu m$.~It is noteworthy that, based on the atomic number density, the blockade spheres for the $|r_{\xi}\rangle$ state, with a radius of 2.0 $\mu m$, are calculated to contain only 7-8 atoms ($n_{sa}=7.5$). If a detuning compensation of $\langle\delta_{s_{\xi}}\rangle\simeq n_{sa}\times 1.0$ GHz is provided, it precisely enables the blockade of the $|s_{\xi}\rangle$ state against the $|r_{\xi}\rangle$ state, whereas the blockade effect among $|r_{\xi}\rangle$ states can be disregarded.

\begin{table}[hbp]
  \centering
  \caption{Diffraction mode under $non$-Hermitian/Hermitian  spatial modulation and $Rydberg$-$switching$ control with detuning phases $\psi_{lc}=0$ and $\psi_{rc}=\psi_{lc}+\Delta\psi$ for left-/right circlar polarization state,  $lc$ ($\circlearrowleft$) and $rc$ ($\circlearrowright$).  }
  \begin{tabular}{c|c|c|c|c}
  \hline\hline
   Phase Difference &  \multicolumn{4}{c} {Diffraction Mode [Asymmetric/Symmetric, A/S]   } \\
   \cline{2-5}
  \multirow{2}{*}{($\Delta\psi=\psi_{rc}-\psi_{lc}$)}  &  \multicolumn{2}{c|} {Switch  on} &  \multicolumn{2}{c} {Switch off} \\
  \cline{2-5}
    & $lc$ ($\circlearrowleft$) & $rc$ ($\circlearrowright$) & $lc$ ($\circlearrowleft$)  & $rc$ ($\circlearrowright$) \\
  \hline\hline
  $\Delta\psi=0$ & $left$ (A) & $left$ (A) & none  & none \\
  $\Delta\psi=\pi/2$ & $left$ (A) & $normal$ (S) &   none  & $Damman$ (S) \\
  $\Delta\psi=\pi$ & $left$ (A) & $right$ (A) &   none & none \\
  \hline\hline
  \end{tabular}
    \label{tab2}
  \end{table}

Based on these principle outlined above, we extend the spatial modulation of the atomic ensemble to a two-dimensional scenario, as $\Omega_{d_{\xi}}(x,y)=\Omega_{d_{0}}[\cos(2\pi x/a)+\cos(2\pi y/a)]$ (or $\delta_{\xi}(x,y)=\delta_{\xi_0}[\sin(4\pi x/a+\psi_{x_{\xi}})+\sin(4\pi y/a+\psi_{y_{\xi}})]$). We further explore the diffraction properties of photons with varying polarizations at different optical depths ($\mathcal{L}=45, 65\mu$m), illustrated in Figure~\ref{Fig6}. Showing in the left four panels Fig.~\ref{Fig6}(a$_1$) to (b$_2$), we achieve a controlled photon polarization beam splitter. Observing the right four ones, Fig.~\ref{Fig6}(c$_1$) to (d$_2$), gate pulses merely demonstrate directional control of diffraction for left-circlar polarized photons, but implement a control mechanism similar to a \textit{odd-Dammann} grating for the optical routing switch of beam splitting for right-circlar polarized photons. In contrast, Fig.~\ref{Fig6}(a$_2$) to (d$_2$) demonstrate that before the application of the gate pulses, there is an intensity distribution present solely in the $0$th order. This indicates that the incident beam undergoes no diffraction, regardless of its polarization state. This methodology underscores the nuanced manipulation of photon paths and splitting strategies, predicated on their polarization characteristics. We have tabulated the achievable diffraction modes and conditions of the atomic grating in Table.~\ref{tab2}.

\section{CONCLUSIONS}\label{PartIV}
In summary, the ultra-cold atomic ensemble driven into a dual-$\mathcal{N}$-type configuration (like ``VVV'') can provide an interesting venue to realize \textit{polarization dependent} non-Hermitian EIG by Rydberg dynamical control. The transition selection rule ensures that the channels of photons with different polarizations have no crosstalk and can be controlled separately by different non-Hermitian space modulation (detuning), which can result in a variety of polarization-selective diffraction modes or functions, such as controlled polarization beam splitting. In addition, spatial tuning mode of grating has been shifted from amplitude spatial modulation to detuning spatial modulation, by the Rydberg cross-blockade effect for both non-crosswalk polarized light transition channels. Therefore, pulse excitation of different gate Rydberg states provides dynamic switching control for corresponding polarization photon diffraction modes.

Through the integration of non-Hermitian optical modulation and based on non-interfering channels, this grating facilitates tunable asymmetric diffraction that correlates with the photons' polarization states. Furthermore, controlled Rydberg excitation provides a non-local control mode as switching effects combined with the optical scattering system. This investigation gives a novel idea for harnessing the polarization degree of freedom of photons within EIG structures. It advances the application of asymmetric optical scattering, underpinned by non-Hermitian optical modulation. The potential demonstrated by Rydberg atomic gratings opens up exciting possibilities for further exploration and utilization in non-Hermitian/ non-local optical manipulation research.

\section*{Acknowledgment}
This work is supported by the National Natural Science Foundation of China (12104107); Jilin Scientific and Technological Development Program (20220101009JC); Scientific Research Project of Jilin Provincial Department of Education (JJKH20241411KJ); Fundamental Research Funds for the Central Universities (2412022ZD046).

\appendix

\section{$\mathcal{APT}$ and Asymmetric Diffraction}\label{SS}
\setcounter{equation}{0}
\renewcommand{\theequation}{A\arabic{equation}}

In this section, we decompose the susceptibility of the medium into its real and imaginary components based on the results presented in Eq.~\eqref{rho_1}, aiming to examine their spatial variation and parity, which is crucial for ascertaining whether they conform to non-Hermitian spatial modulation. Consequently, the following expression is derived:
\begin{align}\label{A1}
\varrho_{\xi}^{\prime}(\omega,x) & =\frac{\gamma\gamma^2_{m_{\xi}}\delta_{\xi}(x)}{[\gamma\gamma_{m_{\xi}}+\Omega_c^2(x)]^2+\gamma^2_{m_{\xi}}\delta^2_{\xi}(x)},\\
\varrho_{\xi}^{\prime\prime}(\omega,x) & =\frac{i[\gamma^2\gamma^2_{m_{\xi}}+\gamma\gamma_{m_{\xi}}\Omega_c^2(x)]}{[\gamma\gamma_{m_{\xi}}+\Omega_c^2(x)]^2+\gamma^2_{m_{\xi}}\delta_{\xi}^2(x)},\nonumber
\end{align}
where $\delta_{\xi}(x)=\delta_{0,\xi}\cdot\sin[2\pi\lambda_c(x-x_0)/a+\psi_{\xi}]$ is an odd function of $x$ with $\psi_{\xi}=2n\pi$ ($n\in \mathbb{Z}$). Obviously, $\varrho_{\xi}^{\prime}(\omega,x) $ and $\varrho_{\xi}^{\prime\prime}(\omega,x)$ are the \textit{odd} and \textit{even} function versus $x$ here, which is satisfied with the optical $\mathcal{PT}$ anti-symmetry condition ($\mathcal{APT}$) with $\chi_{\xi}^{\prime}(x)\propto \varrho_{\xi}^{\prime}(\omega,x)$ ($\chi_{\xi}^{\prime\prime}(x)\propto \varrho_{\xi}^{\prime\prime}(\omega,x)$) and $\chi_{\xi}^{\prime}(-x)=-\chi_{\xi}^{\prime}(x)$ ($\chi_{\xi}^{\prime\prime}(-x)=\chi_{\xi}^{\prime\prime}(x)$).

Focus on the $n$th-order ($n\neq 0$) diffraction by examining $\mathcal{E}_{\xi,n}=\mathcal{E}^{\mathcal{L}}_{\xi}(\theta_n)$. For simplicity, with $\alpha_{\xi}(x)=k_p\int_0^{\mathcal{L}}  \chi_{\xi}^{\prime}(x,z)dz$ $\propto\varrho_{\xi}^{\prime}(\omega,x)$, $\beta_{\xi}(x)=k_p\int_0^{\mathcal{L}} \chi_{\xi}^{\prime\prime}(x,z)dz$ $\propto\varrho_{\xi}^{\prime\prime}(\omega,x)$, and $\gamma_{n}(x)=2n\pi x$, we can make a power series expansion of Eq.~\eqref{Eq_Ip},
\begin{align}
 \mathcal{E}_{\xi,n} & =\int^{\frac{+a}{2}}_{\frac{-a}{2}} e^{-i\gamma_{n}(x)}\sum_0^{m}\frac{[i\alpha_{\xi}(x)]}{m!}^{m}e^{-\beta_{\xi}(x)}dx,
\end{align}
if $\alpha_{\xi}\ll 1$ with small optical depths (OD, $\eta=\frac{8\pi\rho_0\wp_{eg}^2L}{\varepsilon\hbar\lambda_p\gamma}\simeq 11.7$), with $\{m,k\}\in \mathbb{N}$. Defining
 \begin{align}\label{Eq_fg}
 f_{n,\xi}^{\prime} & =-\int^{+a/2}_{-a/2}dx\cdot\alpha_{\xi}(x)\sin[\gamma_n(x)],\\
 f_{n,\xi}^{\prime\prime} & =\int^{+a/2}_{-a/2}dx\cdot\alpha_{\xi}(x)\cos[\gamma_n(x)],\nonumber\\
 g_{n,\xi}^{\prime} & =\int^{+a/2}_{-a/2}dx\cdot\alpha_{\xi}(x)^2\cos[\gamma_n(x)],\nonumber\\
 g_{n,\xi}^{\prime\prime} & =\int^{+a/2}_{-a/2}dx\cdot\alpha_{\xi}(x)^2\sin[\gamma_n(x)],\nonumber
 \end{align}
 with the replacement $\alpha_{\xi}(x)\to \varepsilon_n\alpha_{\xi}(x)$ and low absorption $\beta_{\xi}\to 0$, we further get
 \begin{align}\label{Eq_En}
 \mathcal{E}_n\simeq [f_n^{\prime}\varepsilon_n-g_n^{\prime}\varepsilon_n^2/2]+i[f_n^{\prime\prime}\varepsilon_n-g_n^{\prime\prime}\varepsilon_n^2/2],
 \end{align}
 the scattering factor $\varepsilon_n$ is small enough to keep only the first- and second-order scattering terms \cite{AHEIG,LoopEIG}.
 It is easy to find that $f_n^{\prime}=-f_{-n}^{\prime}$, $f_n^{\prime\prime}=f_{-n}^{\prime\prime}$, $g_n^{\prime}=g_{-n}^{\prime}$ and $g_n^{\prime\prime}=-g_{-n}^{\prime\prime}$, and we can write down the intensities $I_{\pm n}\simeq |f_n^{\prime}\varepsilon_n\pm g_n^{\prime}\varepsilon_n^2/2|^2+|f_n^{\prime\prime}\varepsilon_n\pm g_n^{\prime\prime}\varepsilon_n^2/2|^2$ for the $\pm n$th diffraction orders.  Accordingly, the intensity contrast ratio can be introduced as
 \begin{equation}\label{eta}
 \eta_n=\left\vert\frac{I_n-I_{-n}}{I_n+I_{-n}}\right\vert\simeq 2\left\vert\frac{f_n^{\prime}\cdot g_n^{\prime}+f_n^{\prime\prime}\cdot g_n^{\prime\prime}}{(f_n^{\prime})^2+(f_n^{\prime\prime})^2}\right\vert,	
 \end{equation}
  to evaluate the degree of asymmetric diffraction.~From Eq.~\eqref{A1}, we can get $\alpha_{\xi}(-x)=-\alpha_{\xi}(x)$, and $\gamma_{n}(-x)=-\gamma_{n}(x)$, under the $\mathcal{APT}$ modulation. It is easily obtained that $f_n^{\prime\prime} = g_n^{\prime\prime}=f_n^{\prime\prime}\cdot g_n^{\prime\prime}=0$. After simplification, we can attain the relationship between the asymmetry coefficients and the scattering coefficients
\begin{align}\label{Eq_eta_apt}
\eta_{n,\xi}^{\mathcal{APT}} = 2\left|\frac{ g_{n,\xi}^{\prime}}{f_{n,\xi}^{\prime}}\right| \neq 0,
\end{align}
indicting the absence of high-order scattering terms results in asymmetric diffraction of the grating under $\mathcal{APT}$-symmteric modulation.

\bibliographystyle{apsrev4-1}
\bibliography{Manu.bib}

\begin{thebibliography}{71}%
\makeatletter
\providecommand \@ifxundefined [1]{%
 \@ifx{#1\undefined}
}%
\providecommand \@ifnum [1]{%
 \ifnum #1\expandafter \@firstoftwo
 \else \expandafter \@secondoftwo
 \fi
}%
\providecommand \@ifx [1]{%
 \ifx #1\expandafter \@firstoftwo
 \else \expandafter \@secondoftwo
 \fi
}%
\providecommand \natexlab [1]{#1}%
\providecommand \enquote  [1]{``#1''}%
\providecommand \bibnamefont  [1]{#1}%
\providecommand \bibfnamefont [1]{#1}%
\providecommand \citenamefont [1]{#1}%
\providecommand \href@noop [0]{\@secondoftwo}%
\providecommand \href [0]{\begingroup \@sanitize@url \@href}%
\providecommand \@href[1]{\@@startlink{#1}\@@href}%
\providecommand \@@href[1]{\endgroup#1\@@endlink}%
\providecommand \@sanitize@url [0]{\catcode `\\12\catcode `\$12\catcode `\&12\catcode `\#12\catcode `\^12\catcode `\_12\catcode `\%12\relax}%
\providecommand \@@startlink[1]{}%
\providecommand \@@endlink[0]{}%
\providecommand \url  [0]{\begingroup\@sanitize@url \@url }%
\providecommand \@url [1]{\endgroup\@href {#1}{\urlprefix }}%
\providecommand \urlprefix  [0]{URL }%
\providecommand \Eprint [0]{\href }%
\providecommand \doibase [0]{http://dx.doi.org/}%
\providecommand \selectlanguage [0]{\@gobble}%
\providecommand \bibinfo  [0]{\@secondoftwo}%
\providecommand \bibfield  [0]{\@secondoftwo}%
\providecommand \translation [1]{[#1]}%
\providecommand \BibitemOpen [0]{}%
\providecommand \bibitemStop [0]{}%
\providecommand \bibitemNoStop [0]{.\EOS\space}%
\providecommand \EOS [0]{\spacefactor3000\relax}%
\providecommand \BibitemShut  [1]{\csname bibitem#1\endcsname}%
\let\auto@bib@innerbib\@empty
\bibitem [{\citenamefont {Curtis}\ \emph {et~al.}(2021)\citenamefont {Curtis}, \citenamefont {Hann}, \citenamefont {Elder}, \citenamefont {Wang}, \citenamefont {Frunzio}, \citenamefont {Jiang},\ and\ \citenamefont {Schoelkopf}}]{J2021}%
  \BibitemOpen
  \bibfield  {author} {\bibinfo {author} {\bibfnamefont {J.~C.}\ \bibnamefont {Curtis}}, \bibinfo {author} {\bibfnamefont {C.~T.}\ \bibnamefont {Hann}}, \bibinfo {author} {\bibfnamefont {S.~S.}\ \bibnamefont {Elder}}, \bibinfo {author} {\bibfnamefont {C.~S.}\ \bibnamefont {Wang}}, \bibinfo {author} {\bibfnamefont {L.}~\bibnamefont {Frunzio}}, \bibinfo {author} {\bibfnamefont {L.}~\bibnamefont {Jiang}}, \ and\ \bibinfo {author} {\bibfnamefont {R.~J.}\ \bibnamefont {Schoelkopf}},\ }\href {\doibase 10.1103/PhysRevA.103.023705} {\bibfield  {journal} {\bibinfo  {journal} {Phys. Rev. A}\ }\textbf {\bibinfo {volume} {103}},\ \bibinfo {pages} {023705} (\bibinfo {year} {2021})}\BibitemShut {NoStop}%
\bibitem [{\citenamefont {Delaney}\ \emph {et~al.}(2022)\citenamefont {Delaney}, \citenamefont {Seshadreesan}, \citenamefont {MacCormack}, \citenamefont {Galda}, \citenamefont {Guha},\ and\ \citenamefont {Narang}}]{D2022}%
  \BibitemOpen
  \bibfield  {author} {\bibinfo {author} {\bibfnamefont {C.}~\bibnamefont {Delaney}}, \bibinfo {author} {\bibfnamefont {K.~P.}\ \bibnamefont {Seshadreesan}}, \bibinfo {author} {\bibfnamefont {I.}~\bibnamefont {MacCormack}}, \bibinfo {author} {\bibfnamefont {A.}~\bibnamefont {Galda}}, \bibinfo {author} {\bibfnamefont {S.}~\bibnamefont {Guha}}, \ and\ \bibinfo {author} {\bibfnamefont {P.}~\bibnamefont {Narang}},\ }\href {\doibase 10.1103/PhysRevA.106.032613} {\bibfield  {journal} {\bibinfo  {journal} {Phys. Rev. A}\ }\textbf {\bibinfo {volume} {106}},\ \bibinfo {pages} {032613} (\bibinfo {year} {2022})}\BibitemShut {NoStop}%
\bibitem [{\citenamefont {Wu}\ \emph {et~al.}(2024)\citenamefont {Wu}, \citenamefont {Huang}, \citenamefont {Ji}, \citenamefont {Wang},\ and\ \citenamefont {Chang-Hasnain}}]{C2024}%
  \BibitemOpen
  \bibfield  {author} {\bibinfo {author} {\bibfnamefont {C.}~\bibnamefont {Wu}}, \bibinfo {author} {\bibfnamefont {X.}~\bibnamefont {Huang}}, \bibinfo {author} {\bibfnamefont {Y.}~\bibnamefont {Ji}}, \bibinfo {author} {\bibfnamefont {J.}~\bibnamefont {Wang}}, \ and\ \bibinfo {author} {\bibfnamefont {C.~J.}\ \bibnamefont {Chang-Hasnain}},\ }\href {\doibase 10.1109/JPHOT.2024.3367298} {\bibfield  {journal} {\bibinfo  {journal} {IEEE Photonics J.}\ }\textbf {\bibinfo {volume} {16}},\ \bibinfo {pages} {1} (\bibinfo {year} {2024})}\BibitemShut {NoStop}%
\bibitem [{\citenamefont {Kawakubo}\ and\ \citenamefont {Yamamoto}(2010)}]{OD2010-01}%
  \BibitemOpen
  \bibfield  {author} {\bibinfo {author} {\bibfnamefont {T.}~\bibnamefont {Kawakubo}}\ and\ \bibinfo {author} {\bibfnamefont {K.}~\bibnamefont {Yamamoto}},\ }\href {\doibase 10.1103/PhysRevA.82.032102} {\bibfield  {journal} {\bibinfo  {journal} {Phys. Rev. A}\ }\textbf {\bibinfo {volume} {82}},\ \bibinfo {pages} {032102} (\bibinfo {year} {2010})}\BibitemShut {NoStop}%
\bibitem [{\citenamefont {Tay}\ \emph {et~al.}(2009)\citenamefont {Tay}, \citenamefont {Hsu},\ and\ \citenamefont {Bowen}}]{OD2009-02}%
  \BibitemOpen
  \bibfield  {author} {\bibinfo {author} {\bibfnamefont {J.~W.}\ \bibnamefont {Tay}}, \bibinfo {author} {\bibfnamefont {M.~T.~L.}\ \bibnamefont {Hsu}}, \ and\ \bibinfo {author} {\bibfnamefont {W.~P.}\ \bibnamefont {Bowen}},\ }\href {\doibase 10.1103/PhysRevA.80.063806} {\bibfield  {journal} {\bibinfo  {journal} {Phys. Rev. A}\ }\textbf {\bibinfo {volume} {80}},\ \bibinfo {pages} {063806} (\bibinfo {year} {2009})}\BibitemShut {NoStop}%
\bibitem [{\citenamefont {Then}\ \emph {et~al.}(2014)\citenamefont {Then}, \citenamefont {Razinskas}, \citenamefont {Feichtner}, \citenamefont {Haas}, \citenamefont {Wild}, \citenamefont {Bellini}, \citenamefont {Osellame}, \citenamefont {Cerullo},\ and\ \citenamefont {Hecht}}]{OD2014-03}%
  \BibitemOpen
  \bibfield  {author} {\bibinfo {author} {\bibfnamefont {P.}~\bibnamefont {Then}}, \bibinfo {author} {\bibfnamefont {G.}~\bibnamefont {Razinskas}}, \bibinfo {author} {\bibfnamefont {T.}~\bibnamefont {Feichtner}}, \bibinfo {author} {\bibfnamefont {P.}~\bibnamefont {Haas}}, \bibinfo {author} {\bibfnamefont {A.}~\bibnamefont {Wild}}, \bibinfo {author} {\bibfnamefont {N.}~\bibnamefont {Bellini}}, \bibinfo {author} {\bibfnamefont {R.}~\bibnamefont {Osellame}}, \bibinfo {author} {\bibfnamefont {G.}~\bibnamefont {Cerullo}}, \ and\ \bibinfo {author} {\bibfnamefont {B.}~\bibnamefont {Hecht}},\ }\href {\doibase 10.1103/PhysRevA.89.053801} {\bibfield  {journal} {\bibinfo  {journal} {Phys. Rev. A}\ }\textbf {\bibinfo {volume} {89}},\ \bibinfo {pages} {053801} (\bibinfo {year} {2014})}\BibitemShut {NoStop}%
\bibitem [{\citenamefont {Graf}\ \emph {et~al.}(2005)\citenamefont {Graf}, \citenamefont {Kimball}, \citenamefont {Rochester}, \citenamefont {Kerner}, \citenamefont {Wong}, \citenamefont {Budker}, \citenamefont {Alexandrov}, \citenamefont {Balabas},\ and\ \citenamefont {Yashchuk}}]{OD2005-04}%
  \BibitemOpen
  \bibfield  {author} {\bibinfo {author} {\bibfnamefont {M.~T.}\ \bibnamefont {Graf}}, \bibinfo {author} {\bibfnamefont {D.~F.}\ \bibnamefont {Kimball}}, \bibinfo {author} {\bibfnamefont {S.~M.}\ \bibnamefont {Rochester}}, \bibinfo {author} {\bibfnamefont {K.}~\bibnamefont {Kerner}}, \bibinfo {author} {\bibfnamefont {C.}~\bibnamefont {Wong}}, \bibinfo {author} {\bibfnamefont {D.}~\bibnamefont {Budker}}, \bibinfo {author} {\bibfnamefont {E.~B.}\ \bibnamefont {Alexandrov}}, \bibinfo {author} {\bibfnamefont {M.~V.}\ \bibnamefont {Balabas}}, \ and\ \bibinfo {author} {\bibfnamefont {V.~V.}\ \bibnamefont {Yashchuk}},\ }\href {\doibase 10.1103/PhysRevA.72.023401} {\bibfield  {journal} {\bibinfo  {journal} {Phys. Rev. A}\ }\textbf {\bibinfo {volume} {72}},\ \bibinfo {pages} {023401} (\bibinfo {year} {2005})}\BibitemShut {NoStop}%
\bibitem [{\citenamefont {LaBelle}\ \emph {et~al.}(1996)\citenamefont {LaBelle}, \citenamefont {Hansen}, \citenamefont {Mankowski},\ and\ \citenamefont {Fairbank}}]{OD1996-05}%
  \BibitemOpen
  \bibfield  {author} {\bibinfo {author} {\bibfnamefont {R.~D.}\ \bibnamefont {LaBelle}}, \bibinfo {author} {\bibfnamefont {C.~S.}\ \bibnamefont {Hansen}}, \bibinfo {author} {\bibfnamefont {M.~M.}\ \bibnamefont {Mankowski}}, \ and\ \bibinfo {author} {\bibfnamefont {W.~M.}\ \bibnamefont {Fairbank}},\ }\href {\doibase 10.1103/PhysRevA.54.4461} {\bibfield  {journal} {\bibinfo  {journal} {Phys. Rev. A}\ }\textbf {\bibinfo {volume} {54}},\ \bibinfo {pages} {4461} (\bibinfo {year} {1996})}\BibitemShut {NoStop}%
\bibitem [{\citenamefont {Selig}\ \emph {et~al.}(2015)\citenamefont {Selig}, \citenamefont {Siffels},\ and\ \citenamefont {Rezus}}]{S2015-01}%
  \BibitemOpen
  \bibfield  {author} {\bibinfo {author} {\bibfnamefont {O.}~\bibnamefont {Selig}}, \bibinfo {author} {\bibfnamefont {R.}~\bibnamefont {Siffels}}, \ and\ \bibinfo {author} {\bibfnamefont {Y.~L.~A.}\ \bibnamefont {Rezus}},\ }\href {\doibase 10.1103/PhysRevLett.114.233004} {\bibfield  {journal} {\bibinfo  {journal} {Phys. Rev. Lett.}\ }\textbf {\bibinfo {volume} {114}},\ \bibinfo {pages} {233004} (\bibinfo {year} {2015})}\BibitemShut {NoStop}%
\bibitem [{\citenamefont {Coddington}\ \emph {et~al.}(2008)\citenamefont {Coddington}, \citenamefont {Swann},\ and\ \citenamefont {Newbury}}]{S2008-02}%
  \BibitemOpen
  \bibfield  {author} {\bibinfo {author} {\bibfnamefont {I.}~\bibnamefont {Coddington}}, \bibinfo {author} {\bibfnamefont {W.~C.}\ \bibnamefont {Swann}}, \ and\ \bibinfo {author} {\bibfnamefont {N.~R.}\ \bibnamefont {Newbury}},\ }\href {\doibase 10.1103/PhysRevLett.100.013902} {\bibfield  {journal} {\bibinfo  {journal} {Phys. Rev. Lett.}\ }\textbf {\bibinfo {volume} {100}},\ \bibinfo {pages} {013902} (\bibinfo {year} {2008})}\BibitemShut {NoStop}%
\bibitem [{\citenamefont {Agarwal}\ and\ \citenamefont {Scully}(1996)}]{S1996-03}%
  \BibitemOpen
  \bibfield  {author} {\bibinfo {author} {\bibfnamefont {G.~S.}\ \bibnamefont {Agarwal}}\ and\ \bibinfo {author} {\bibfnamefont {M.~O.}\ \bibnamefont {Scully}},\ }\href {\doibase 10.1103/PhysRevA.53.467} {\bibfield  {journal} {\bibinfo  {journal} {Phys. Rev. A}\ }\textbf {\bibinfo {volume} {53}},\ \bibinfo {pages} {467} (\bibinfo {year} {1996})}\BibitemShut {NoStop}%
\bibitem [{\citenamefont {Sinclair}\ \emph {et~al.}(2011)\citenamefont {Sinclair}, \citenamefont {Cossel}, \citenamefont {Coffey}, \citenamefont {Ye},\ and\ \citenamefont {Cornell}}]{S2011-04}%
  \BibitemOpen
  \bibfield  {author} {\bibinfo {author} {\bibfnamefont {L.~C.}\ \bibnamefont {Sinclair}}, \bibinfo {author} {\bibfnamefont {K.~C.}\ \bibnamefont {Cossel}}, \bibinfo {author} {\bibfnamefont {T.}~\bibnamefont {Coffey}}, \bibinfo {author} {\bibfnamefont {J.}~\bibnamefont {Ye}}, \ and\ \bibinfo {author} {\bibfnamefont {E.~A.}\ \bibnamefont {Cornell}},\ }\href {\doibase 10.1103/PhysRevLett.107.093002} {\bibfield  {journal} {\bibinfo  {journal} {Phys. Rev. Lett.}\ }\textbf {\bibinfo {volume} {107}},\ \bibinfo {pages} {093002} (\bibinfo {year} {2011})}\BibitemShut {NoStop}%
\bibitem [{\citenamefont {Zhao}\ and\ \citenamefont {Wright}(1999)}]{S1999-05}%
  \BibitemOpen
  \bibfield  {author} {\bibinfo {author} {\bibfnamefont {W.}~\bibnamefont {Zhao}}\ and\ \bibinfo {author} {\bibfnamefont {J.~C.}\ \bibnamefont {Wright}},\ }\href {\doibase 10.1103/PhysRevLett.83.1950} {\bibfield  {journal} {\bibinfo  {journal} {Phys. Rev. Lett.}\ }\textbf {\bibinfo {volume} {83}},\ \bibinfo {pages} {1950} (\bibinfo {year} {1999})}\BibitemShut {NoStop}%
\bibitem [{\citenamefont {Reuter}\ \emph {et~al.}(1997)\citenamefont {Reuter}, \citenamefont {Bernhardt}, \citenamefont {Wedler}, \citenamefont {Schardt}, \citenamefont {Starke},\ and\ \citenamefont {Heinz}}]{HI1997}%
  \BibitemOpen
  \bibfield  {author} {\bibinfo {author} {\bibfnamefont {K.}~\bibnamefont {Reuter}}, \bibinfo {author} {\bibfnamefont {J.}~\bibnamefont {Bernhardt}}, \bibinfo {author} {\bibfnamefont {H.}~\bibnamefont {Wedler}}, \bibinfo {author} {\bibfnamefont {J.}~\bibnamefont {Schardt}}, \bibinfo {author} {\bibfnamefont {U.}~\bibnamefont {Starke}}, \ and\ \bibinfo {author} {\bibfnamefont {K.}~\bibnamefont {Heinz}},\ }\href {\doibase 10.1103/PhysRevLett.79.4818} {\bibfield  {journal} {\bibinfo  {journal} {Phys. Rev. Lett.}\ }\textbf {\bibinfo {volume} {79}},\ \bibinfo {pages} {4818} (\bibinfo {year} {1997})}\BibitemShut {NoStop}%
\bibitem [{\citenamefont {Asghar}\ \emph {et~al.}(2019)\citenamefont {Asghar}, \citenamefont {Abbas}, \citenamefont {Qamar},\ and\ \citenamefont {Qamar}}]{AS2019}%
  \BibitemOpen
  \bibfield  {author} {\bibinfo {author} {\bibfnamefont {S.}~\bibnamefont {Asghar}}, \bibinfo {author} {\bibfnamefont {M.}~\bibnamefont {Abbas}}, \bibinfo {author} {\bibfnamefont {S.}~\bibnamefont {Qamar}}, \ and\ \bibinfo {author} {\bibfnamefont {S.}~\bibnamefont {Qamar}},\ }\href {\doibase https://doi.org/10.1016/j.optcom.2018.12.056} {\bibfield  {journal} {\bibinfo  {journal} {Opt. Commun.}\ }\textbf {\bibinfo {volume} {437}},\ \bibinfo {pages} {290} (\bibinfo {year} {2019})}\BibitemShut {NoStop}%
\bibitem [{\citenamefont {Ransom}(1980)}]{PL1980}%
  \BibitemOpen
  \bibfield  {author} {\bibinfo {author} {\bibfnamefont {P.~L.}\ \bibnamefont {Ransom}},\ }\href {\doibase 10.1364/OL.5.000327} {\bibfield  {journal} {\bibinfo  {journal} {Opt. Lett.}\ }\textbf {\bibinfo {volume} {5}},\ \bibinfo {pages} {327} (\bibinfo {year} {1980})}\BibitemShut {NoStop}%
\bibitem [{\citenamefont {Latychevskaia}\ and\ \citenamefont {Fink}(2007)}]{L2007}%
  \BibitemOpen
  \bibfield  {author} {\bibinfo {author} {\bibfnamefont {T.}~\bibnamefont {Latychevskaia}}\ and\ \bibinfo {author} {\bibfnamefont {H.-W.}\ \bibnamefont {Fink}},\ }\href {\doibase 10.1103/PhysRevLett.98.233901} {\bibfield  {journal} {\bibinfo  {journal} {Phys. Rev. Lett.}\ }\textbf {\bibinfo {volume} {98}},\ \bibinfo {pages} {233901} (\bibinfo {year} {2007})}\BibitemShut {NoStop}%
\bibitem [{\citenamefont {Cser}\ \emph {et~al.}(2002)\citenamefont {Cser}, \citenamefont {T\"or\"ok}, \citenamefont {Krexner}, \citenamefont {Sharkov},\ and\ \citenamefont {Farag\'o}}]{CL2002}%
  \BibitemOpen
  \bibfield  {author} {\bibinfo {author} {\bibfnamefont {L.}~\bibnamefont {Cser}}, \bibinfo {author} {\bibfnamefont {G.}~\bibnamefont {T\"or\"ok}}, \bibinfo {author} {\bibfnamefont {G.}~\bibnamefont {Krexner}}, \bibinfo {author} {\bibfnamefont {I.}~\bibnamefont {Sharkov}}, \ and\ \bibinfo {author} {\bibfnamefont {B.}~\bibnamefont {Farag\'o}},\ }\href {\doibase 10.1103/PhysRevLett.89.175504} {\bibfield  {journal} {\bibinfo  {journal} {Phys. Rev. Lett.}\ }\textbf {\bibinfo {volume} {89}},\ \bibinfo {pages} {175504} (\bibinfo {year} {2002})}\BibitemShut {NoStop}%
\bibitem [{\citenamefont {Galeotti}\ \emph {et~al.}(2010)\citenamefont {Galeotti}, \citenamefont {Siegel},\ and\ \citenamefont {Stetten}}]{JM2010}%
  \BibitemOpen
  \bibfield  {author} {\bibinfo {author} {\bibfnamefont {J.~M.}\ \bibnamefont {Galeotti}}, \bibinfo {author} {\bibfnamefont {M.}~\bibnamefont {Siegel}}, \ and\ \bibinfo {author} {\bibfnamefont {G.}~\bibnamefont {Stetten}},\ }\href {\doibase 10.1364/OL.35.002352} {\bibfield  {journal} {\bibinfo  {journal} {Opt. Lett.}\ }\textbf {\bibinfo {volume} {35}},\ \bibinfo {pages} {2352} (\bibinfo {year} {2010})}\BibitemShut {NoStop}%
\bibitem [{\citenamefont {Hong}\ \emph {et~al.}(2014)\citenamefont {Hong}, \citenamefont {Yeom}, \citenamefont {Jang}, \citenamefont {Hong},\ and\ \citenamefont {Lee}}]{HK2014}%
  \BibitemOpen
  \bibfield  {author} {\bibinfo {author} {\bibfnamefont {K.}~\bibnamefont {Hong}}, \bibinfo {author} {\bibfnamefont {J.}~\bibnamefont {Yeom}}, \bibinfo {author} {\bibfnamefont {C.}~\bibnamefont {Jang}}, \bibinfo {author} {\bibfnamefont {J.}~\bibnamefont {Hong}}, \ and\ \bibinfo {author} {\bibfnamefont {B.}~\bibnamefont {Lee}},\ }\href {\doibase 10.1364/OL.39.000127} {\bibfield  {journal} {\bibinfo  {journal} {Opt. Lett.}\ }\textbf {\bibinfo {volume} {39}},\ \bibinfo {pages} {127} (\bibinfo {year} {2014})}\BibitemShut {NoStop}%
\bibitem [{\citenamefont {Markman}\ \emph {et~al.}(2016)\citenamefont {Markman}, \citenamefont {Shen}, \citenamefont {Hua},\ and\ \citenamefont {Javidi}}]{AM2016}%
  \BibitemOpen
  \bibfield  {author} {\bibinfo {author} {\bibfnamefont {A.}~\bibnamefont {Markman}}, \bibinfo {author} {\bibfnamefont {X.}~\bibnamefont {Shen}}, \bibinfo {author} {\bibfnamefont {H.}~\bibnamefont {Hua}}, \ and\ \bibinfo {author} {\bibfnamefont {B.}~\bibnamefont {Javidi}},\ }\href {\doibase 10.1364/OL.41.000297} {\bibfield  {journal} {\bibinfo  {journal} {Opt. Lett.}\ }\textbf {\bibinfo {volume} {41}},\ \bibinfo {pages} {297} (\bibinfo {year} {2016})}\BibitemShut {NoStop}%
\bibitem [{\citenamefont {Braun}\ \emph {et~al.}(1997)\citenamefont {Braun}, \citenamefont {Kane},\ and\ \citenamefont {Norris}}]{AB1997}%
  \BibitemOpen
  \bibfield  {author} {\bibinfo {author} {\bibfnamefont {A.}~\bibnamefont {Braun}}, \bibinfo {author} {\bibfnamefont {S.}~\bibnamefont {Kane}}, \ and\ \bibinfo {author} {\bibfnamefont {T.}~\bibnamefont {Norris}},\ }\href {\doibase 10.1364/OL.22.000615} {\bibfield  {journal} {\bibinfo  {journal} {Opt. Lett.}\ }\textbf {\bibinfo {volume} {22}},\ \bibinfo {pages} {615} (\bibinfo {year} {1997})}\BibitemShut {NoStop}%
\bibitem [{\citenamefont {Galvanauskas}\ \emph {et~al.}(1998)\citenamefont {Galvanauskas}, \citenamefont {Harter}, \citenamefont {Arbore}, \citenamefont {Chou},\ and\ \citenamefont {Fejer}}]{AG1998}%
  \BibitemOpen
  \bibfield  {author} {\bibinfo {author} {\bibfnamefont {A.}~\bibnamefont {Galvanauskas}}, \bibinfo {author} {\bibfnamefont {D.}~\bibnamefont {Harter}}, \bibinfo {author} {\bibfnamefont {M.~A.}\ \bibnamefont {Arbore}}, \bibinfo {author} {\bibfnamefont {M.~H.}\ \bibnamefont {Chou}}, \ and\ \bibinfo {author} {\bibfnamefont {M.~M.}\ \bibnamefont {Fejer}},\ }\href {\doibase 10.1364/OL.23.001695} {\bibfield  {journal} {\bibinfo  {journal} {Opt. Lett.}\ }\textbf {\bibinfo {volume} {23}},\ \bibinfo {pages} {1695} (\bibinfo {year} {1998})}\BibitemShut {NoStop}%
\bibitem [{\citenamefont {Forget}\ \emph {et~al.}(2005)\citenamefont {Forget}, \citenamefont {Cotel}, \citenamefont {Brambrink}, \citenamefont {Audebert}, \citenamefont {Blanc}, \citenamefont {Jullien}, \citenamefont {Albert},\ and\ \citenamefont {Ch\'{e}riaux}}]{NF2005}%
  \BibitemOpen
  \bibfield  {author} {\bibinfo {author} {\bibfnamefont {N.}~\bibnamefont {Forget}}, \bibinfo {author} {\bibfnamefont {A.}~\bibnamefont {Cotel}}, \bibinfo {author} {\bibfnamefont {E.}~\bibnamefont {Brambrink}}, \bibinfo {author} {\bibfnamefont {P.}~\bibnamefont {Audebert}}, \bibinfo {author} {\bibfnamefont {C.~L.}\ \bibnamefont {Blanc}}, \bibinfo {author} {\bibfnamefont {A.}~\bibnamefont {Jullien}}, \bibinfo {author} {\bibfnamefont {O.}~\bibnamefont {Albert}}, \ and\ \bibinfo {author} {\bibfnamefont {G.}~\bibnamefont {Ch\'{e}riaux}},\ }\href {\doibase 10.1364/OL.30.002921} {\bibfield  {journal} {\bibinfo  {journal} {Opt. Lett.}\ }\textbf {\bibinfo {volume} {30}},\ \bibinfo {pages} {2921} (\bibinfo {year} {2005})}\BibitemShut {NoStop}%
\bibitem [{\citenamefont {Kim}\ \emph {et~al.}(2003)\citenamefont {Kim}, \citenamefont {Lee}, \citenamefont {Hong}, \citenamefont {Kim}, \citenamefont {Choi},\ and\ \citenamefont {Nam}}]{HT2003}%
  \BibitemOpen
  \bibfield  {author} {\bibinfo {author} {\bibfnamefont {H.~T.}\ \bibnamefont {Kim}}, \bibinfo {author} {\bibfnamefont {D.~G.}\ \bibnamefont {Lee}}, \bibinfo {author} {\bibfnamefont {K.-H.}\ \bibnamefont {Hong}}, \bibinfo {author} {\bibfnamefont {J.-H.}\ \bibnamefont {Kim}}, \bibinfo {author} {\bibfnamefont {I.~W.}\ \bibnamefont {Choi}}, \ and\ \bibinfo {author} {\bibfnamefont {C.~H.}\ \bibnamefont {Nam}},\ }\href {\doibase 10.1103/PhysRevA.67.051801} {\bibfield  {journal} {\bibinfo  {journal} {Phys. Rev. A}\ }\textbf {\bibinfo {volume} {67}},\ \bibinfo {pages} {051801} (\bibinfo {year} {2003})}\BibitemShut {NoStop}%
\bibitem [{\citenamefont {Shen}\ \emph {et~al.}(2017)\citenamefont {Shen}, \citenamefont {Gao}, \citenamefont {Meng}, \citenamefont {Fu},\ and\ \citenamefont {Gong}}]{YS2017}%
  \BibitemOpen
  \bibfield  {author} {\bibinfo {author} {\bibfnamefont {Y.}~\bibnamefont {Shen}}, \bibinfo {author} {\bibfnamefont {G.}~\bibnamefont {Gao}}, \bibinfo {author} {\bibfnamefont {Y.}~\bibnamefont {Meng}}, \bibinfo {author} {\bibfnamefont {X.}~\bibnamefont {Fu}}, \ and\ \bibinfo {author} {\bibfnamefont {M.}~\bibnamefont {Gong}},\ }\href {\doibase 10.1103/PhysRevA.96.043851} {\bibfield  {journal} {\bibinfo  {journal} {Phys. Rev. A}\ }\textbf {\bibinfo {volume} {96}},\ \bibinfo {pages} {043851} (\bibinfo {year} {2017})}\BibitemShut {NoStop}%
\bibitem [{\citenamefont {Ling}\ \emph {et~al.}(1998)\citenamefont {Ling}, \citenamefont {Li},\ and\ \citenamefont {Xiao}}]{H1998}%
  \BibitemOpen
  \bibfield  {author} {\bibinfo {author} {\bibfnamefont {H.~Y.}\ \bibnamefont {Ling}}, \bibinfo {author} {\bibfnamefont {Y.-Q.}\ \bibnamefont {Li}}, \ and\ \bibinfo {author} {\bibfnamefont {M.}~\bibnamefont {Xiao}},\ }\href {\doibase 10.1103/PhysRevA.57.1338} {\bibfield  {journal} {\bibinfo  {journal} {Phys. Rev. A}\ }\textbf {\bibinfo {volume} {57}},\ \bibinfo {pages} {1338} (\bibinfo {year} {1998})}\BibitemShut {NoStop}%
\bibitem [{\citenamefont {Mitsunaga}\ and\ \citenamefont {Imoto}(1999)}]{M1999}%
  \BibitemOpen
  \bibfield  {author} {\bibinfo {author} {\bibfnamefont {M.}~\bibnamefont {Mitsunaga}}\ and\ \bibinfo {author} {\bibfnamefont {N.}~\bibnamefont {Imoto}},\ }\href {\doibase 10.1103/PhysRevA.59.4773} {\bibfield  {journal} {\bibinfo  {journal} {Phys. Rev. A}\ }\textbf {\bibinfo {volume} {59}},\ \bibinfo {pages} {4773} (\bibinfo {year} {1999})}\BibitemShut {NoStop}%
\bibitem [{\citenamefont {Hang}\ \emph {et~al.}(2019)\citenamefont {Hang}, \citenamefont {Li},\ and\ \citenamefont {Huang}}]{HC2019}%
  \BibitemOpen
  \bibfield  {author} {\bibinfo {author} {\bibfnamefont {C.}~\bibnamefont {Hang}}, \bibinfo {author} {\bibfnamefont {W.}~\bibnamefont {Li}}, \ and\ \bibinfo {author} {\bibfnamefont {G.}~\bibnamefont {Huang}},\ }\href {\doibase 10.1103/PhysRevA.100.043807} {\bibfield  {journal} {\bibinfo  {journal} {Phys. Rev. A}\ }\textbf {\bibinfo {volume} {100}},\ \bibinfo {pages} {043807} (\bibinfo {year} {2019})}\BibitemShut {NoStop}%
\bibitem [{\citenamefont {Gao}\ \emph {et~al.}(2022)\citenamefont {Gao}, \citenamefont {Hang},\ and\ \citenamefont {Huang}}]{G2022}%
  \BibitemOpen
  \bibfield  {author} {\bibinfo {author} {\bibfnamefont {J.}~\bibnamefont {Gao}}, \bibinfo {author} {\bibfnamefont {C.}~\bibnamefont {Hang}}, \ and\ \bibinfo {author} {\bibfnamefont {G.}~\bibnamefont {Huang}},\ }\href {\doibase 10.1103/PhysRevA.105.063511} {\bibfield  {journal} {\bibinfo  {journal} {Phys. Rev. A}\ }\textbf {\bibinfo {volume} {105}},\ \bibinfo {pages} {063511} (\bibinfo {year} {2022})}\BibitemShut {NoStop}%
\bibitem [{\citenamefont {Harris}\ \emph {et~al.}(1990)\citenamefont {Harris}, \citenamefont {Field},\ and\ \citenamefont {Imamo\ifmmode~\breve{g}\else \u{g}\fi{}lu}}]{H1990}%
  \BibitemOpen
  \bibfield  {author} {\bibinfo {author} {\bibfnamefont {S.~E.}\ \bibnamefont {Harris}}, \bibinfo {author} {\bibfnamefont {J.~E.}\ \bibnamefont {Field}}, \ and\ \bibinfo {author} {\bibfnamefont {A.}~\bibnamefont {Imamo\ifmmode~\breve{g}\else \u{g}\fi{}lu}},\ }\href {\doibase 10.1103/PhysRevLett.64.1107} {\bibfield  {journal} {\bibinfo  {journal} {Phys. Rev. Lett.}\ }\textbf {\bibinfo {volume} {64}},\ \bibinfo {pages} {1107} (\bibinfo {year} {1990})}\BibitemShut {NoStop}%
\bibitem [{\citenamefont {Harris}(1997)}]{H1997}%
  \BibitemOpen
  \bibfield  {author} {\bibinfo {author} {\bibfnamefont {S.~E.}\ \bibnamefont {Harris}},\ }\href {\doibase 10.1063/1.881806} {\bibfield  {journal} {\bibinfo  {journal} {Phys. Today}\ }\textbf {\bibinfo {volume} {50}},\ \bibinfo {pages} {36} (\bibinfo {year} {1997})}\BibitemShut {NoStop}%
\bibitem [{\citenamefont {Ou}\ and\ \citenamefont {Huang}(2024)}]{O2024}%
  \BibitemOpen
  \bibfield  {author} {\bibinfo {author} {\bibfnamefont {Y.}~\bibnamefont {Ou}}\ and\ \bibinfo {author} {\bibfnamefont {G.}~\bibnamefont {Huang}},\ }\href {\doibase 10.1103/PhysRevA.109.023508} {\bibfield  {journal} {\bibinfo  {journal} {Phys. Rev. A}\ }\textbf {\bibinfo {volume} {109}},\ \bibinfo {pages} {023508} (\bibinfo {year} {2024})}\BibitemShut {NoStop}%
\bibitem [{\citenamefont {Zhou}\ \emph {et~al.}(2013)\citenamefont {Zhou}, \citenamefont {Qi}, \citenamefont {Sun}, \citenamefont {Chen}, \citenamefont {Yang}, \citenamefont {Niu},\ and\ \citenamefont {Gong}}]{ZFX2013}%
  \BibitemOpen
  \bibfield  {author} {\bibinfo {author} {\bibfnamefont {F.}~\bibnamefont {Zhou}}, \bibinfo {author} {\bibfnamefont {Y.}~\bibnamefont {Qi}}, \bibinfo {author} {\bibfnamefont {H.}~\bibnamefont {Sun}}, \bibinfo {author} {\bibfnamefont {D.}~\bibnamefont {Chen}}, \bibinfo {author} {\bibfnamefont {J.}~\bibnamefont {Yang}}, \bibinfo {author} {\bibfnamefont {Y.}~\bibnamefont {Niu}}, \ and\ \bibinfo {author} {\bibfnamefont {S.}~\bibnamefont {Gong}},\ }\href {\doibase 10.1364/OE.21.012249} {\bibfield  {journal} {\bibinfo  {journal} {Opt. Express}\ }\textbf {\bibinfo {volume} {21}},\ \bibinfo {pages} {12249} (\bibinfo {year} {2013})}\BibitemShut {NoStop}%
\bibitem [{\citenamefont {Dong}\ \emph {et~al.}(2017)\citenamefont {Dong}, \citenamefont {Li},\ and\ \citenamefont {Zhou}}]{DY2017}%
  \BibitemOpen
  \bibfield  {author} {\bibinfo {author} {\bibfnamefont {Y.-B.}\ \bibnamefont {Dong}}, \bibinfo {author} {\bibfnamefont {J.-Y.}\ \bibnamefont {Li}}, \ and\ \bibinfo {author} {\bibfnamefont {Z.-Y.}\ \bibnamefont {Zhou}},\ }\href {\doibase 10.1088/1674-1056/26/1/014202} {\bibfield  {journal} {\bibinfo  {journal} {Chin. Phys. B}\ }\textbf {\bibinfo {volume} {26}},\ \bibinfo {pages} {014202} (\bibinfo {year} {2017})}\BibitemShut {NoStop}%
\bibitem [{\citenamefont {You}\ \emph {et~al.}(2019)\citenamefont {You}, \citenamefont {Qi}, \citenamefont {Niu},\ and\ \citenamefont {Gong}}]{YY2019}%
  \BibitemOpen
  \bibfield  {author} {\bibinfo {author} {\bibfnamefont {Y.}~\bibnamefont {You}}, \bibinfo {author} {\bibfnamefont {Y.-H.}\ \bibnamefont {Qi}}, \bibinfo {author} {\bibfnamefont {Y.-P.}\ \bibnamefont {Niu}}, \ and\ \bibinfo {author} {\bibfnamefont {S.-Q.}\ \bibnamefont {Gong}},\ }\href {\doibase 10.1088/1361-648X/aaf8c3} {\bibfield  {journal} {\bibinfo  {journal} {J. Phys.: Condens.Matter}\ }\textbf {\bibinfo {volume} {31}},\ \bibinfo {pages} {105801} (\bibinfo {year} {2019})}\BibitemShut {NoStop}%
\bibitem [{\citenamefont {Carvalho}\ and\ \citenamefont {de~Araujo}(2011)}]{CS2011}%
  \BibitemOpen
  \bibfield  {author} {\bibinfo {author} {\bibfnamefont {S.~A.}\ \bibnamefont {Carvalho}}\ and\ \bibinfo {author} {\bibfnamefont {L.~E.~E.}\ \bibnamefont {de~Araujo}},\ }\href {\doibase 10.1364/OE.19.001936} {\bibfield  {journal} {\bibinfo  {journal} {Opt. Express}\ }\textbf {\bibinfo {volume} {19}},\ \bibinfo {pages} {1936} (\bibinfo {year} {2011})}\BibitemShut {NoStop}%
\bibitem [{\citenamefont {Vafafard}\ and\ \citenamefont {Mahmoudi}(2015)}]{VA2015}%
  \BibitemOpen
  \bibfield  {author} {\bibinfo {author} {\bibfnamefont {A.}~\bibnamefont {Vafafard}}\ and\ \bibinfo {author} {\bibfnamefont {M.}~\bibnamefont {Mahmoudi}},\ }\href {\doibase 10.1364/AO.54.010613} {\bibfield  {journal} {\bibinfo  {journal} {Appl. Opt.}\ }\textbf {\bibinfo {volume} {54}},\ \bibinfo {pages} {10613} (\bibinfo {year} {2015})}\BibitemShut {NoStop}%
\bibitem [{\citenamefont {Hongju}\ \emph {et~al.}(2018)\citenamefont {Hongju}, \citenamefont {Bin}, \citenamefont {Yihong},\ and\ \citenamefont {Yandong}}]{GHJ2018}%
  \BibitemOpen
  \bibfield  {author} {\bibinfo {author} {\bibfnamefont {G.}~\bibnamefont {Hongju}}, \bibinfo {author} {\bibfnamefont {C.}~\bibnamefont {Bin}}, \bibinfo {author} {\bibfnamefont {Q.}~\bibnamefont {Yihong}}, \ and\ \bibinfo {author} {\bibfnamefont {P.}~\bibnamefont {Yandong}},\ }\href {\doibase 10.1080/09500340.2017.1410587} {\bibfield  {journal} {\bibinfo  {journal} {J. Mod. Opt.}\ }\textbf {\bibinfo {volume} {65}},\ \bibinfo {pages} {852} (\bibinfo {year} {2018})}\BibitemShut {NoStop}%
\bibitem [{\citenamefont {Meng}\ \emph {et~al.}(2021{\natexlab{a}})\citenamefont {Meng}, \citenamefont {Liu}, \citenamefont {Liu}, \citenamefont {Feng},\ and\ \citenamefont {Qiu}}]{MDZ2021}%
  \BibitemOpen
  \bibfield  {author} {\bibinfo {author} {\bibfnamefont {D.}~\bibnamefont {Meng}}, \bibinfo {author} {\bibfnamefont {M.}~\bibnamefont {Liu}}, \bibinfo {author} {\bibfnamefont {Y.}~\bibnamefont {Liu}}, \bibinfo {author} {\bibfnamefont {Y.}~\bibnamefont {Feng}}, \ and\ \bibinfo {author} {\bibfnamefont {T.}~\bibnamefont {Qiu}},\ }\href {\doibase 10.1007/s10773-021-04914-w} {\bibfield  {journal} {\bibinfo  {journal} {Int. J. Theor. Phys.}\ }\textbf {\bibinfo {volume} {60}},\ \bibinfo {pages} {3387} (\bibinfo {year} {2021}{\natexlab{a}})}\BibitemShut {NoStop}%
\bibitem [{\citenamefont {El-Ganainy}\ \emph {et~al.}(2007)\citenamefont {El-Ganainy}, \citenamefont {Makris}, \citenamefont {Christodoulides},\ and\ \citenamefont {Musslimani}}]{REG2007}%
  \BibitemOpen
  \bibfield  {author} {\bibinfo {author} {\bibfnamefont {R.}~\bibnamefont {El-Ganainy}}, \bibinfo {author} {\bibfnamefont {K.~G.}\ \bibnamefont {Makris}}, \bibinfo {author} {\bibfnamefont {D.~N.}\ \bibnamefont {Christodoulides}}, \ and\ \bibinfo {author} {\bibfnamefont {Z.~H.}\ \bibnamefont {Musslimani}},\ }\href {\doibase 10.1364/OL.32.002632} {\bibfield  {journal} {\bibinfo  {journal} {Opt. Lett.}\ }\textbf {\bibinfo {volume} {32}},\ \bibinfo {pages} {2632} (\bibinfo {year} {2007})}\BibitemShut {NoStop}%
\bibitem [{\citenamefont {Bender}(2007)}]{B2007}%
  \BibitemOpen
  \bibfield  {author} {\bibinfo {author} {\bibfnamefont {C.~M.}\ \bibnamefont {Bender}},\ }\href {\doibase 10.1088/0034-4885/70/6/R03} {\bibfield  {journal} {\bibinfo  {journal} {Rep. Prog. Phys.}\ }\textbf {\bibinfo {volume} {70}},\ \bibinfo {pages} {947} (\bibinfo {year} {2007})}\BibitemShut {NoStop}%
\bibitem [{\citenamefont {Ge}\ and\ \citenamefont {T\"ureci}(2013)}]{GL2013}%
  \BibitemOpen
  \bibfield  {author} {\bibinfo {author} {\bibfnamefont {L.}~\bibnamefont {Ge}}\ and\ \bibinfo {author} {\bibfnamefont {H.~E.}\ \bibnamefont {T\"ureci}},\ }\href {\doibase 10.1103/PhysRevA.88.053810} {\bibfield  {journal} {\bibinfo  {journal} {Phys. Rev. A}\ }\textbf {\bibinfo {volume} {88}},\ \bibinfo {pages} {053810} (\bibinfo {year} {2013})}\BibitemShut {NoStop}%
\bibitem [{\citenamefont {Ma}\ \emph {et~al.}(2019{\natexlab{a}})\citenamefont {Ma}, \citenamefont {Yu}, \citenamefont {Zhao},\ and\ \citenamefont {Qian}}]{MDD2019}%
  \BibitemOpen
  \bibfield  {author} {\bibinfo {author} {\bibfnamefont {D.}~\bibnamefont {Ma}}, \bibinfo {author} {\bibfnamefont {D.}~\bibnamefont {Yu}}, \bibinfo {author} {\bibfnamefont {X.-D.}\ \bibnamefont {Zhao}}, \ and\ \bibinfo {author} {\bibfnamefont {J.}~\bibnamefont {Qian}},\ }\href {\doibase 10.1103/PhysRevA.99.033826} {\bibfield  {journal} {\bibinfo  {journal} {Phys. Rev. A}\ }\textbf {\bibinfo {volume} {99}},\ \bibinfo {pages} {033826} (\bibinfo {year} {2019}{\natexlab{a}})}\BibitemShut {NoStop}%
\bibitem [{\citenamefont {Yuan}\ \emph {et~al.}(2020)\citenamefont {Yuan}, \citenamefont {Dong}, \citenamefont {Wu}, \citenamefont {Wang}, \citenamefont {Xiao},\ and\ \citenamefont {Jia}}]{YJP2020}%
  \BibitemOpen
  \bibfield  {author} {\bibinfo {author} {\bibfnamefont {J.}~\bibnamefont {Yuan}}, \bibinfo {author} {\bibfnamefont {S.}~\bibnamefont {Dong}}, \bibinfo {author} {\bibfnamefont {C.}~\bibnamefont {Wu}}, \bibinfo {author} {\bibfnamefont {L.}~\bibnamefont {Wang}}, \bibinfo {author} {\bibfnamefont {L.}~\bibnamefont {Xiao}}, \ and\ \bibinfo {author} {\bibfnamefont {S.}~\bibnamefont {Jia}},\ }\href {\doibase 10.1364/OE.400618} {\bibfield  {journal} {\bibinfo  {journal} {Opt. Express}\ }\textbf {\bibinfo {volume} {28}},\ \bibinfo {pages} {23820} (\bibinfo {year} {2020})}\BibitemShut {NoStop}%
\bibitem [{\citenamefont {Wu}\ \emph {et~al.}(2014)\citenamefont {Wu}, \citenamefont {Artoni},\ and\ \citenamefont {La~Rocca}}]{JHW2014}%
  \BibitemOpen
  \bibfield  {author} {\bibinfo {author} {\bibfnamefont {J.-H.}\ \bibnamefont {Wu}}, \bibinfo {author} {\bibfnamefont {M.}~\bibnamefont {Artoni}}, \ and\ \bibinfo {author} {\bibfnamefont {G.~C.}\ \bibnamefont {La~Rocca}},\ }\href {\doibase 10.1103/PhysRevLett.113.123004} {\bibfield  {journal} {\bibinfo  {journal} {Phys. Rev. Lett.}\ }\textbf {\bibinfo {volume} {113}},\ \bibinfo {pages} {123004} (\bibinfo {year} {2014})}\BibitemShut {NoStop}%
\bibitem [{\citenamefont {Peng}\ \emph {et~al.}(2016)\citenamefont {Peng}, \citenamefont {Cao}, \citenamefont {Shen}, \citenamefont {Qu}, \citenamefont {Wen}, \citenamefont {Jiang},\ and\ \citenamefont {Xiao}}]{PP2016}%
  \BibitemOpen
  \bibfield  {author} {\bibinfo {author} {\bibfnamefont {P.}~\bibnamefont {Peng}}, \bibinfo {author} {\bibfnamefont {W.}~\bibnamefont {Cao}}, \bibinfo {author} {\bibfnamefont {C.}~\bibnamefont {Shen}}, \bibinfo {author} {\bibfnamefont {W.}~\bibnamefont {Qu}}, \bibinfo {author} {\bibfnamefont {J.}~\bibnamefont {Wen}}, \bibinfo {author} {\bibfnamefont {L.}~\bibnamefont {Jiang}}, \ and\ \bibinfo {author} {\bibfnamefont {Y.}~\bibnamefont {Xiao}},\ }\href {\doibase 10.1038/nphys3842} {\bibfield  {journal} {\bibinfo  {journal} {Nat. Phys.}\ }\textbf {\bibinfo {volume} {12}},\ \bibinfo {pages} {1139} (\bibinfo {year} {2016})}\BibitemShut {NoStop}%
\bibitem [{\citenamefont {Yang}\ \emph {et~al.}(2017)\citenamefont {Yang}, \citenamefont {Liu},\ and\ \citenamefont {You}}]{YF2017}%
  \BibitemOpen
  \bibfield  {author} {\bibinfo {author} {\bibfnamefont {F.}~\bibnamefont {Yang}}, \bibinfo {author} {\bibfnamefont {Y.-C.}\ \bibnamefont {Liu}}, \ and\ \bibinfo {author} {\bibfnamefont {L.}~\bibnamefont {You}},\ }\href {\doibase 10.1103/PhysRevA.96.053845} {\bibfield  {journal} {\bibinfo  {journal} {Phys. Rev. A}\ }\textbf {\bibinfo {volume} {96}},\ \bibinfo {pages} {053845} (\bibinfo {year} {2017})}\BibitemShut {NoStop}%
\bibitem [{\citenamefont {Asghar}\ \emph {et~al.}(2016)\citenamefont {Asghar}, \citenamefont {Ziauddin}, \citenamefont {Qamar},\ and\ \citenamefont {Qamar}}]{SA2016}%
  \BibitemOpen
  \bibfield  {author} {\bibinfo {author} {\bibfnamefont {S.}~\bibnamefont {Asghar}}, \bibinfo {author} {\bibnamefont {Ziauddin}}, \bibinfo {author} {\bibfnamefont {S.}~\bibnamefont {Qamar}}, \ and\ \bibinfo {author} {\bibfnamefont {S.}~\bibnamefont {Qamar}},\ }\href {\doibase 10.1103/PhysRevA.94.033823} {\bibfield  {journal} {\bibinfo  {journal} {Phys. Rev. A}\ }\textbf {\bibinfo {volume} {94}},\ \bibinfo {pages} {033823} (\bibinfo {year} {2016})}\BibitemShut {NoStop}%
\bibitem [{\citenamefont {Liu}\ \emph {et~al.}(2016)\citenamefont {Liu}, \citenamefont {Tian}, \citenamefont {Wang}, \citenamefont {Yan},\ and\ \citenamefont {Wu}}]{LYM2016}%
  \BibitemOpen
  \bibfield  {author} {\bibinfo {author} {\bibfnamefont {Y.-M.}\ \bibnamefont {Liu}}, \bibinfo {author} {\bibfnamefont {X.-D.}\ \bibnamefont {Tian}}, \bibinfo {author} {\bibfnamefont {X.}~\bibnamefont {Wang}}, \bibinfo {author} {\bibfnamefont {D.}~\bibnamefont {Yan}}, \ and\ \bibinfo {author} {\bibfnamefont {J.-H.}\ \bibnamefont {Wu}},\ }\href {\doibase 10.1364/OL.41.000408} {\bibfield  {journal} {\bibinfo  {journal} {Opt. Lett.}\ }\textbf {\bibinfo {volume} {41}},\ \bibinfo {pages} {408} (\bibinfo {year} {2016})}\BibitemShut {NoStop}%
\bibitem [{\citenamefont {Bozorgzadeh}\ and\ \citenamefont {Sahrai}(2018)}]{BF2018}%
  \BibitemOpen
  \bibfield  {author} {\bibinfo {author} {\bibfnamefont {F.}~\bibnamefont {Bozorgzadeh}}\ and\ \bibinfo {author} {\bibfnamefont {M.}~\bibnamefont {Sahrai}},\ }\href {\doibase 10.1103/PhysRevA.98.043822} {\bibfield  {journal} {\bibinfo  {journal} {Phys. Rev. A}\ }\textbf {\bibinfo {volume} {98}},\ \bibinfo {pages} {043822} (\bibinfo {year} {2018})}\BibitemShut {NoStop}%
\bibitem [{\citenamefont {Ma}\ \emph {et~al.}(2019{\natexlab{b}})\citenamefont {Ma}, \citenamefont {Yu}, \citenamefont {Zhao},\ and\ \citenamefont {Qian}}]{MD2019}%
  \BibitemOpen
  \bibfield  {author} {\bibinfo {author} {\bibfnamefont {D.}~\bibnamefont {Ma}}, \bibinfo {author} {\bibfnamefont {D.}~\bibnamefont {Yu}}, \bibinfo {author} {\bibfnamefont {X.-D.}\ \bibnamefont {Zhao}}, \ and\ \bibinfo {author} {\bibfnamefont {J.}~\bibnamefont {Qian}},\ }\href {\doibase 10.1103/PhysRevA.99.033826} {\bibfield  {journal} {\bibinfo  {journal} {Phys. Rev. A}\ }\textbf {\bibinfo {volume} {99}},\ \bibinfo {pages} {033826} (\bibinfo {year} {2019}{\natexlab{b}})}\BibitemShut {NoStop}%
\bibitem [{\citenamefont {Barreiro}\ and\ \citenamefont {Tabosa}(2003)}]{BS2003}%
  \BibitemOpen
  \bibfield  {author} {\bibinfo {author} {\bibfnamefont {S.}~\bibnamefont {Barreiro}}\ and\ \bibinfo {author} {\bibfnamefont {J.~W.~R.}\ \bibnamefont {Tabosa}},\ }\href {\doibase 10.1103/PhysRevLett.90.133001} {\bibfield  {journal} {\bibinfo  {journal} {Phys. Rev. Lett.}\ }\textbf {\bibinfo {volume} {90}},\ \bibinfo {pages} {133001} (\bibinfo {year} {2003})}\BibitemShut {NoStop}%
\bibitem [{\citenamefont {Asadpour}\ \emph {et~al.}(2022)\citenamefont {Asadpour}, \citenamefont {Hamedi}, \citenamefont {Kirova},\ and\ \citenamefont {Paspalakis}}]{AS2022}%
  \BibitemOpen
  \bibfield  {author} {\bibinfo {author} {\bibfnamefont {S.~H.}\ \bibnamefont {Asadpour}}, \bibinfo {author} {\bibfnamefont {H.~R.}\ \bibnamefont {Hamedi}}, \bibinfo {author} {\bibfnamefont {T.}~\bibnamefont {Kirova}}, \ and\ \bibinfo {author} {\bibfnamefont {E.}~\bibnamefont {Paspalakis}},\ }\href {\doibase 10.1103/PhysRevA.105.043709} {\bibfield  {journal} {\bibinfo  {journal} {Phys. Rev. A}\ }\textbf {\bibinfo {volume} {105}},\ \bibinfo {pages} {043709} (\bibinfo {year} {2022})}\BibitemShut {NoStop}%
\bibitem [{\citenamefont {Jinlan}\ \emph {et~al.}(2022)\citenamefont {Jinlan}, \citenamefont {Shuifa},\ and\ \citenamefont {Dongbiao}}]{JJL2022}%
  \BibitemOpen
  \bibfield  {author} {\bibinfo {author} {\bibfnamefont {J.}~\bibnamefont {Jinlan}}, \bibinfo {author} {\bibfnamefont {S.}~\bibnamefont {Shuifa}}, \ and\ \bibinfo {author} {\bibfnamefont {K.}~\bibnamefont {Dongbiao}},\ }\href {\doibase 10.1088/1612-202X/ac50a6} {\bibfield  {journal} {\bibinfo  {journal} {Laser Phys. Lett.}\ }\textbf {\bibinfo {volume} {19}},\ \bibinfo {pages} {045202} (\bibinfo {year} {2022})}\BibitemShut {NoStop}%
\bibitem [{\citenamefont {Tang}\ \emph {et~al.}(2022)\citenamefont {Tang}, \citenamefont {Zeinali},\ and\ \citenamefont {Abdulkareem}}]{TZG2022}%
  \BibitemOpen
  \bibfield  {author} {\bibinfo {author} {\bibfnamefont {Z.}~\bibnamefont {Tang}}, \bibinfo {author} {\bibfnamefont {B.}~\bibnamefont {Zeinali}}, \ and\ \bibinfo {author} {\bibfnamefont {S.~S.}\ \bibnamefont {Abdulkareem}},\ }\href {\doibase 10.1088/1612-202X/ac5e39} {\bibfield  {journal} {\bibinfo  {journal} {Laser Phys. Lett.}\ }\textbf {\bibinfo {volume} {19}},\ \bibinfo {pages} {055204} (\bibinfo {year} {2022})}\BibitemShut {NoStop}%
\bibitem [{\citenamefont {Wahab}\ \emph {et~al.}(2023)\citenamefont {Wahab}, \citenamefont {Abbas},\ and\ \citenamefont {Sanders}}]{WA2023}%
  \BibitemOpen
  \bibfield  {author} {\bibinfo {author} {\bibfnamefont {A.}~\bibnamefont {Wahab}}, \bibinfo {author} {\bibfnamefont {M.}~\bibnamefont {Abbas}}, \ and\ \bibinfo {author} {\bibfnamefont {B.~C.}\ \bibnamefont {Sanders}},\ }\href {\doibase 10.1088/1367-2630/accc6e} {\bibfield  {journal} {\bibinfo  {journal} {New J. Phys.}\ }\textbf {\bibinfo {volume} {25}},\ \bibinfo {pages} {053003} (\bibinfo {year} {2023})}\BibitemShut {NoStop}%
\bibitem [{\citenamefont {He}\ \emph {et~al.}(1998)\citenamefont {He}, \citenamefont {Nose},\ and\ \citenamefont {Sato}}]{ZH1998}%
  \BibitemOpen
  \bibfield  {author} {\bibinfo {author} {\bibfnamefont {Z.}~\bibnamefont {He}}, \bibinfo {author} {\bibfnamefont {T.}~\bibnamefont {Nose}}, \ and\ \bibinfo {author} {\bibfnamefont {S.}~\bibnamefont {Sato}},\ }\href {\doibase 10.1117/1.601876} {\bibfield  {journal} {\bibinfo  {journal} {Opt. Eng.}\ }\textbf {\bibinfo {volume} {37}},\ \bibinfo {pages} {2885 } (\bibinfo {year} {1998})}\BibitemShut {NoStop}%
\bibitem [{\citenamefont {Lu}\ \emph {et~al.}(2004)\citenamefont {Lu}, \citenamefont {Du},\ and\ \citenamefont {Wu}}]{LY2004}%
  \BibitemOpen
  \bibfield  {author} {\bibinfo {author} {\bibfnamefont {Y.-Q.}\ \bibnamefont {Lu}}, \bibinfo {author} {\bibfnamefont {F.}~\bibnamefont {Du}}, \ and\ \bibinfo {author} {\bibfnamefont {S.-T.}\ \bibnamefont {Wu}},\ }\href {\doibase 10.1063/1.1633337} {\bibfield  {journal} {\bibinfo  {journal} {J. Appl. Phys.}\ }\textbf {\bibinfo {volume} {95}},\ \bibinfo {pages} {810} (\bibinfo {year} {2004})}\BibitemShut {NoStop}%
\bibitem [{\citenamefont {An}(2007)}]{A2007}%
  \BibitemOpen
  \bibfield  {author} {\bibinfo {author} {\bibfnamefont {J.-W.}\ \bibnamefont {An}},\ }\href {\doibase 10.1109/LPT.2007.891652} {\bibfield  {journal} {\bibinfo  {journal} {IEEE Photonics Technol. Lett.}\ }\textbf {\bibinfo {volume} {19}},\ \bibinfo {pages} {369} (\bibinfo {year} {2007})}\BibitemShut {NoStop}%
\bibitem [{\citenamefont {Dreven\ifmmode \check{s}\else~\v{s}\fi{}ek Olenik}\ \emph {et~al.}(2006)\citenamefont {Dreven\ifmmode \check{s}\else~\v{s}\fi{}ek Olenik}, \citenamefont {Fally},\ and\ \citenamefont {Ellabban}}]{D2017}%
  \BibitemOpen
  \bibfield  {author} {\bibinfo {author} {\bibfnamefont {I.}~\bibnamefont {Dreven\ifmmode \check{s}\else~\v{s}\fi{}ek Olenik}}, \bibinfo {author} {\bibfnamefont {M.}~\bibnamefont {Fally}}, \ and\ \bibinfo {author} {\bibfnamefont {M.~A.}\ \bibnamefont {Ellabban}},\ }\href {\doibase 10.1103/PhysRevE.74.021707} {\bibfield  {journal} {\bibinfo  {journal} {Phys. Rev. E}\ }\textbf {\bibinfo {volume} {74}},\ \bibinfo {pages} {021707} (\bibinfo {year} {2006})}\BibitemShut {NoStop}%
\bibitem [{\citenamefont {Pors}\ \emph {et~al.}(2015)\citenamefont {Pors}, \citenamefont {Nielsen},\ and\ \citenamefont {Bozhevolnyi}}]{PA2015}%
  \BibitemOpen
  \bibfield  {author} {\bibinfo {author} {\bibfnamefont {A.}~\bibnamefont {Pors}}, \bibinfo {author} {\bibfnamefont {M.~G.}\ \bibnamefont {Nielsen}}, \ and\ \bibinfo {author} {\bibfnamefont {S.~I.}\ \bibnamefont {Bozhevolnyi}},\ }\href {\doibase 10.1364/OPTICA.2.000716} {\bibfield  {journal} {\bibinfo  {journal} {Optica}\ }\textbf {\bibinfo {volume} {2}},\ \bibinfo {pages} {716} (\bibinfo {year} {2015})}\BibitemShut {NoStop}%
\bibitem [{\citenamefont {Arbabi}\ \emph {et~al.}(2015)\citenamefont {Arbabi}, \citenamefont {Horie}, \citenamefont {Bagheri},\ and\ \citenamefont {Faraon}}]{AA2015}%
  \BibitemOpen
  \bibfield  {author} {\bibinfo {author} {\bibfnamefont {A.}~\bibnamefont {Arbabi}}, \bibinfo {author} {\bibfnamefont {Y.}~\bibnamefont {Horie}}, \bibinfo {author} {\bibfnamefont {M.}~\bibnamefont {Bagheri}}, \ and\ \bibinfo {author} {\bibfnamefont {A.}~\bibnamefont {Faraon}},\ }\href {\doibase 10.1038/nnano.2015.186} {\bibfield  {journal} {\bibinfo  {journal} {Nat. Nanotechnol.}\ }\textbf {\bibinfo {volume} {10}},\ \bibinfo {pages} {937} (\bibinfo {year} {2015})}\BibitemShut {NoStop}%
\bibitem [{\citenamefont {Balthasar~Mueller}\ \emph {et~al.}(2017)\citenamefont {Balthasar~Mueller}, \citenamefont {Rubin}, \citenamefont {Devlin}, \citenamefont {Groever},\ and\ \citenamefont {Capasso}}]{MJP2017}%
  \BibitemOpen
  \bibfield  {author} {\bibinfo {author} {\bibfnamefont {J.~P.}\ \bibnamefont {Balthasar~Mueller}}, \bibinfo {author} {\bibfnamefont {N.~A.}\ \bibnamefont {Rubin}}, \bibinfo {author} {\bibfnamefont {R.~C.}\ \bibnamefont {Devlin}}, \bibinfo {author} {\bibfnamefont {B.}~\bibnamefont {Groever}}, \ and\ \bibinfo {author} {\bibfnamefont {F.}~\bibnamefont {Capasso}},\ }\href {\doibase 10.1103/PhysRevLett.118.113901} {\bibfield  {journal} {\bibinfo  {journal} {Phys. Rev. Lett.}\ }\textbf {\bibinfo {volume} {118}},\ \bibinfo {pages} {113901} (\bibinfo {year} {2017})}\BibitemShut {NoStop}%
\bibitem [{\citenamefont {Rubin}\ \emph {et~al.}(2019)\citenamefont {Rubin}, \citenamefont {D'Aversa}, \citenamefont {Chevalier}, \citenamefont {Shi}, \citenamefont {Chen},\ and\ \citenamefont {Capasso}}]{RN2019}%
  \BibitemOpen
  \bibfield  {author} {\bibinfo {author} {\bibfnamefont {N.~A.}\ \bibnamefont {Rubin}}, \bibinfo {author} {\bibfnamefont {G.}~\bibnamefont {D'Aversa}}, \bibinfo {author} {\bibfnamefont {P.}~\bibnamefont {Chevalier}}, \bibinfo {author} {\bibfnamefont {Z.}~\bibnamefont {Shi}}, \bibinfo {author} {\bibfnamefont {W.~T.}\ \bibnamefont {Chen}}, \ and\ \bibinfo {author} {\bibfnamefont {F.}~\bibnamefont {Capasso}},\ }\href {\doibase 10.1126/science.aax1839} {\bibfield  {journal} {\bibinfo  {journal} {Science}\ }\textbf {\bibinfo {volume} {365}},\ \bibinfo {pages} {eaax1839} (\bibinfo {year} {2019})}\BibitemShut {NoStop}%
\bibitem [{\citenamefont {Yan}\ \emph {et~al.}(2023)\citenamefont {Yan}, \citenamefont {Wang}, \citenamefont {Hu}, \citenamefont {Hao},\ and\ \citenamefont {Bian}}]{YR2023}%
  \BibitemOpen
  \bibfield  {author} {\bibinfo {author} {\bibfnamefont {R.}~\bibnamefont {Yan}}, \bibinfo {author} {\bibfnamefont {W.}~\bibnamefont {Wang}}, \bibinfo {author} {\bibfnamefont {Y.}~\bibnamefont {Hu}}, \bibinfo {author} {\bibfnamefont {Q.}~\bibnamefont {Hao}}, \ and\ \bibinfo {author} {\bibfnamefont {L.}~\bibnamefont {Bian}},\ }\href {\doibase 10.3390/nano13091542} {\bibfield  {journal} {\bibinfo  {journal} {Nanomaterials}\ }\textbf {\bibinfo {volume} {13}},\ \bibinfo {pages} {1542} (\bibinfo {year} {2023})}\BibitemShut {NoStop}%
\bibitem [{\citenamefont {Zhao}(2018)}]{EPIG1}%
  \BibitemOpen
  \bibfield  {author} {\bibinfo {author} {\bibfnamefont {L.}~\bibnamefont {Zhao}},\ }\href {\doibase 10.1038/s41598-018-21494-8} {\bibfield  {journal} {\bibinfo  {journal} {Sci. Rep.}\ }\textbf {\bibinfo {volume} {8}},\ \bibinfo {pages} {3073} (\bibinfo {year} {2018})}\BibitemShut {NoStop}%
\bibitem [{\citenamefont {Meng}\ \emph {et~al.}(2021{\natexlab{b}})\citenamefont {Meng}, \citenamefont {Liu}, \citenamefont {Liu}, \citenamefont {Feng},\ and\ \citenamefont {Qiu}}]{EPIG2}%
  \BibitemOpen
  \bibfield  {author} {\bibinfo {author} {\bibfnamefont {D.}~\bibnamefont {Meng}}, \bibinfo {author} {\bibfnamefont {M.}~\bibnamefont {Liu}}, \bibinfo {author} {\bibfnamefont {Y.}~\bibnamefont {Liu}}, \bibinfo {author} {\bibfnamefont {Y.}~\bibnamefont {Feng}}, \ and\ \bibinfo {author} {\bibfnamefont {T.}~\bibnamefont {Qiu}},\ }\href {\doibase 10.1007/s10773-021-04914-w} {\bibfield  {journal} {\bibinfo  {journal} {Int. J. Theo. Phys.}\ }\textbf {\bibinfo {volume} {60}},\ \bibinfo {pages} {3387} (\bibinfo {year} {2021}{\natexlab{b}})}\BibitemShut {NoStop}%
\bibitem [{\citenamefont {Asadpour}\ and\ \citenamefont {Faizabadi}(2022)}]{EPIG3}%
  \BibitemOpen
  \bibfield  {author} {\bibinfo {author} {\bibfnamefont {S.~H.}\ \bibnamefont {Asadpour}}\ and\ \bibinfo {author} {\bibfnamefont {E.}~\bibnamefont {Faizabadi}},\ }\href {\doibase 10.1364/AO.469098} {\bibfield  {journal} {\bibinfo  {journal} {Appl. Opt.}\ }\textbf {\bibinfo {volume} {61}},\ \bibinfo {pages} {8139} (\bibinfo {year} {2022})}\BibitemShut {NoStop}%
\bibitem [{\citenamefont {Huo}\ \emph {et~al.}(2023)\citenamefont {Huo}, \citenamefont {Hua}, \citenamefont {Tian},\ and\ \citenamefont {Liu}}]{LoopEIG}%
  \BibitemOpen
  \bibfield  {author} {\bibinfo {author} {\bibfnamefont {D.}~\bibnamefont {Huo}}, \bibinfo {author} {\bibfnamefont {S.}~\bibnamefont {Hua}}, \bibinfo {author} {\bibfnamefont {X.-D.}\ \bibnamefont {Tian}}, \ and\ \bibinfo {author} {\bibfnamefont {Y.-M.}\ \bibnamefont {Liu}},\ }\href {\doibase 10.1364/OE.483806} {\bibfield  {journal} {\bibinfo  {journal} {Opt. Express}\ }\textbf {\bibinfo {volume} {31}},\ \bibinfo {pages} {16251} (\bibinfo {year} {2023})}\BibitemShut {NoStop}%
\bibitem [{\citenamefont {Hua}\ \emph {et~al.}(2022)\citenamefont {Hua}, \citenamefont {Liu}, \citenamefont {Lio}, \citenamefont {Zhang}, \citenamefont {Wu}, \citenamefont {Artoni},\ and\ \citenamefont {La~Rocca}}]{AHEIG}%
  \BibitemOpen
  \bibfield  {author} {\bibinfo {author} {\bibfnamefont {S.}~\bibnamefont {Hua}}, \bibinfo {author} {\bibfnamefont {Y.-M.}\ \bibnamefont {Liu}}, \bibinfo {author} {\bibfnamefont {G.~E.}\ \bibnamefont {Lio}}, \bibinfo {author} {\bibfnamefont {X.-J.}\ \bibnamefont {Zhang}}, \bibinfo {author} {\bibfnamefont {J.-H.}\ \bibnamefont {Wu}}, \bibinfo {author} {\bibfnamefont {M.}~\bibnamefont {Artoni}}, \ and\ \bibinfo {author} {\bibfnamefont {G.~C.}\ \bibnamefont {La~Rocca}},\ }\href {\doibase 10.1103/PhysRevResearch.4.023113} {\bibfield  {journal} {\bibinfo  {journal} {Phys. Rev. Res.}\ }\textbf {\bibinfo {volume} {4}},\ \bibinfo {pages} {023113} (\bibinfo {year} {2022})}\BibitemShut {NoStop}%
\end{thebibliography}%

\end{document}